# TAILS OF A RECENT COMET

---

## THE ROLE COMETARY JETS PLAY IN CRUSTAL FORMATION

## ESKER / DRUMLIN SWARMS

MILTON ZYSMAN and FRANK WALLACE
Presented at the Society for Interdisciplinary Studies Cambridge Conference:
Natural Catastrophes during Bronze Age Civilizations, 11-13 July 1997

Revised by MILTON ZYSMAN
June 29, 2009
This revision has substituted higher quality illustrations and single
word corrections. There were no substantive changes to the text.

Edited by HEIKO SANTOSH and MARIUS OLTEAN
July 3, 2009

## Table of Contents

# ABSTRACT

Drain away the earth's oceans and a global pattern of great ridges appears. Adjacent to these continental and undersea mountain ranges are layers of silt and clay, so thick that they fill the gaps between ridges, creating extensive plateaus.

Ranging across this planet's higher latitudes are thousands of tiny replicas of these ridge systems. These esker and drumlin swarms run up hills and across streams in roughly parallel discontinuous strands for hundreds of kilometers.

Preserved by encapsulation in the ice and snow of our last ice age, eskers, drumlins and their related structures will be the focus of this paper. We contend that Greater and lesser ridge systems alike, including the water and sediments that fill them are cometary debris.

Each ridge may be traced to a single stream, or 'jet' of disintegrating materials emanating from shifting areas on a comet's nucleus. A band of these jets, captured into planetary orbit, will land its debris in a unique manner. All debris will be laid down in a sheet perpendicular to the planetary surface, this process which results in the sharp ridge profile, is consistent with the manner in which comets discharge debris along the plane of their orbit.

The jet particles, massive enough to resist planetary atmosphere [sands, gravel and boulders] once landed become robust structures through compacting and immediate concretization. The water and lighter materials, diverted by winds and post-depositional melting, flowed and settled into the inter-ridge basins.

All ridges display internal structural anomalies connected with their cometary origin. The systems lie directly upon older strata. The ridges are free of fossils and do not show signs of organization by hydraulic processes. The cements necessary for their immediate conversion to rock could not be provided from earthly sources.

The establishment of the ridge complexes found on earth is therefore consistent with its initial encounter with a great comet. The ancestral body's return in reduced and fragmented form laid down the lesser esker and drumlin formations.

# INTRODUCTION

**Ridges**

The earth's most prominent geological feature is its ridge systems. The greatest examples are our continental and sub-oceanic mountain belts. The most prolific ridge systems are built on a smaller scale and seldom rise more than 50 meters above surrounding terrain. The best defined of these ridges are eskers and drumlins which are found trending roughly parallel to each other in multiple discontinuous strands. Encapsulated by the deep snows of our most recent ice age, their well preserved remains will be the subject of this paper.

Eskers, drumlins and their related structures are the most distinctive product of the *drift* – a general term for all rock material believed to have been transported by glaciers and deposited directly from the ice or through the agency of melt water.

During the latter part of the previous century geologists were certain that all drift originated on the surface of glacial ice. This paper agrees with this simple observation and seeks to demonstrate a clear connection between the drift and cometary debris.

**Common Enigmas**

Whatever their size or location, ridge systems share critical features supporting an extraterrestrial source.

Consider the tendency of most ridges to travel in roughly parallel discontinuous strands - this is such an ordinary feature of our landscape, that one assumes it to be a well understood geological feature. It is not. Ridge swarms, both great and small along with the finer water-borne debris that mantles their sides and fills the adjacent valleys and basins, are belts of exotic terrain, unconnected to older, underlying strata. Recent probes beneath mountain ranges have convinced geologists that mountains rest upon but are not derived from the underlying strata. (1)

Nineteenth century theoreticians, unbound by this discovery, favored the concept that parallel ridge sequences were corrugations pushed up by a cooling of the earth's contracting crust. This idea was abandoned when calculations revealed that not enough local crustal fabric would have been available through sub-surface contraction to completely cover the ridges. Moreover, the sterile, unsorted interiors of mountain ridges did not match the stratified fossil-rich materials in adjacent valleys. A competing theory attempting to deal with the conversion of great embankments of rock debris into mountains by means of heat and pressure, hypothesized cycles of vertical submersions and levitations of huge sedimentary deposits. This theory is being slowly abandoned as a result of the discovery that the foundations of mountains do not pierce through but lie intact upon older strata.

The lesser ridge systems, like their giant counterparts, have a similar surficial disposition, and though there is general agreement that they have been laid down from ice, there is no common agreement on how the debris was generated, moved and built into eskers and drumlins.

In addition to their corresponding external design, all ridge swarms share equally critical internal similarities. Their interiors are devoid of fossils; their debris is unsorted and shows no evidence of stratification or particle grading that would indicate that running water played a role in their lay down. Moreover sometimes an entire ridge or part of its core is cemented into concretions of sand and gravel. The large quantities of cements needed to turn mountains and lesser ridges into concrete are best produced, as will be argued below, in outer space.

**Comets**

Like all important geological phenomena ridge formations, both great and small are assumed to have evolved from the earth's interior. We contend that each ridge may be traced to a single stream of cometary debris called a jet, discharging intermittently from a comets nucleus. A comets dust tail is made up of a series of jets spreading along the comet's orbital plane in a band of ribbons. It is this unusual attribute of comets that accounts for the unique patterns developed by the earth's ridge systems.

It is these ribbons, landing perpendicular to the earth's surface that creates the unmistakable imprint of a ridge system.

We shall propose a hypothetical two-stage sequence explaining the laydown and structure of the greater and lesser ridge systems.

In the first mountain building stage, a great comet would leave part of its tail in an earth orbit. Portions of the orbiting ribbons of debris would then descend to form continental and undersea mountain ranges.

This ancestral comet would then return later in reduced and fragmented form to lay down the earth's lesser ridge swarms. These latter day fragments, each overlapping to form a compound tail, diminished in scale, but much greater in number, changed continental environments radically.

**Why Eskers and Drumlins?**

We have chosen eskers, drumlins and their related formations as our primary investigative target for the following reasons.

First, they are recent - the product of the retreat of the last continental ice sheets, an event that in some cases is only several thousands of years old.

Second, their structures are well preserved by snow and ice. Their size and condition make them easy to survey and excavate. They are widely distributed on most continents, and when not mined for city and highway construction, are easily accessible.

Third, the century old glacial hypothesis, unlike the mountain building theory, which is consistently shifting, presents a stable target.

# THE LESSER JET RIDGES

Eskers, drumlins and their related structures appear to the observer as narrow discontinuous, roughly parallel ridges. These groupings are often referred to as swarms. (2)

Individual swarms appear to be part of a greater radiating pattern, hundreds of kilometers long. These 'super-swarms' appear to emanate from an imaginary centre called a divide. (Illustrations I and 2) It is assumed that great ice domes, such as those still active in Greenland, somehow ground out, transported, and aligned these great ridge systems.

**Swarms**

Perhaps the most puzzling attribute of eskers and drumlins is their propensity to wander over hill and stream while maintaining their formation. Conveying and depositing these characteristic formations have stretched the imaginations of geologists for over a century.

Individual strands of an esker/drumlin complex may track parallel to other members of the swarm, separated by many kilometers. (3) Some drumlin and esker strands however, come close enough to merge and form sediment filled plateaus called compound ridges. (4)

Ridges sometimes touch, ride up and over one another and then diverge. (Illustration 3) Some drumlins actually ride up the backs of their partners. (5)

**Eskers**

Ridges are made up of closely packed aggregates: boulders, gravel and sand. They can rise as high as 80 meters, but seldom exceed 40 meters. (6) (Illustrations 4 and 5)

Few of the segments found in a swarm are more than a few kilometers long, but these may be traced, in file, crossing hills and streams for hundreds of kilometers. (7)

Esker ridges can become narrow enough for partners to shake hands over the ridge line or wide enough to serve as a highway.

To the casual observer, esker ridges when they are not obscured by overgrowth, may look like abandoned railway embankments. (8) Their distinctive, regular shapes have led to local folk lore. In North America, they have been considered the remnants of aboriginal monuments. (9) In northern Europe a tradition persists of goblins spreading eskers through holes in the bottom of giant sandbags. (10)

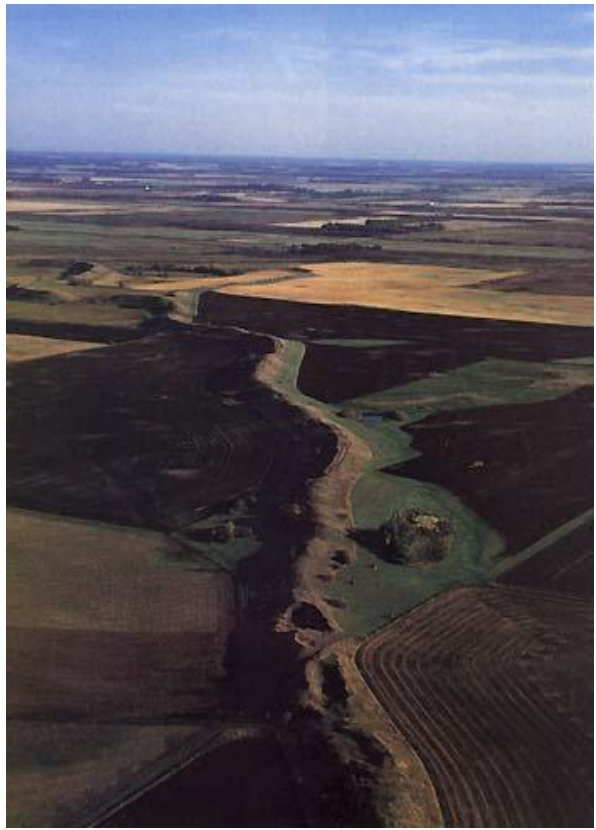

**Illustration 1**

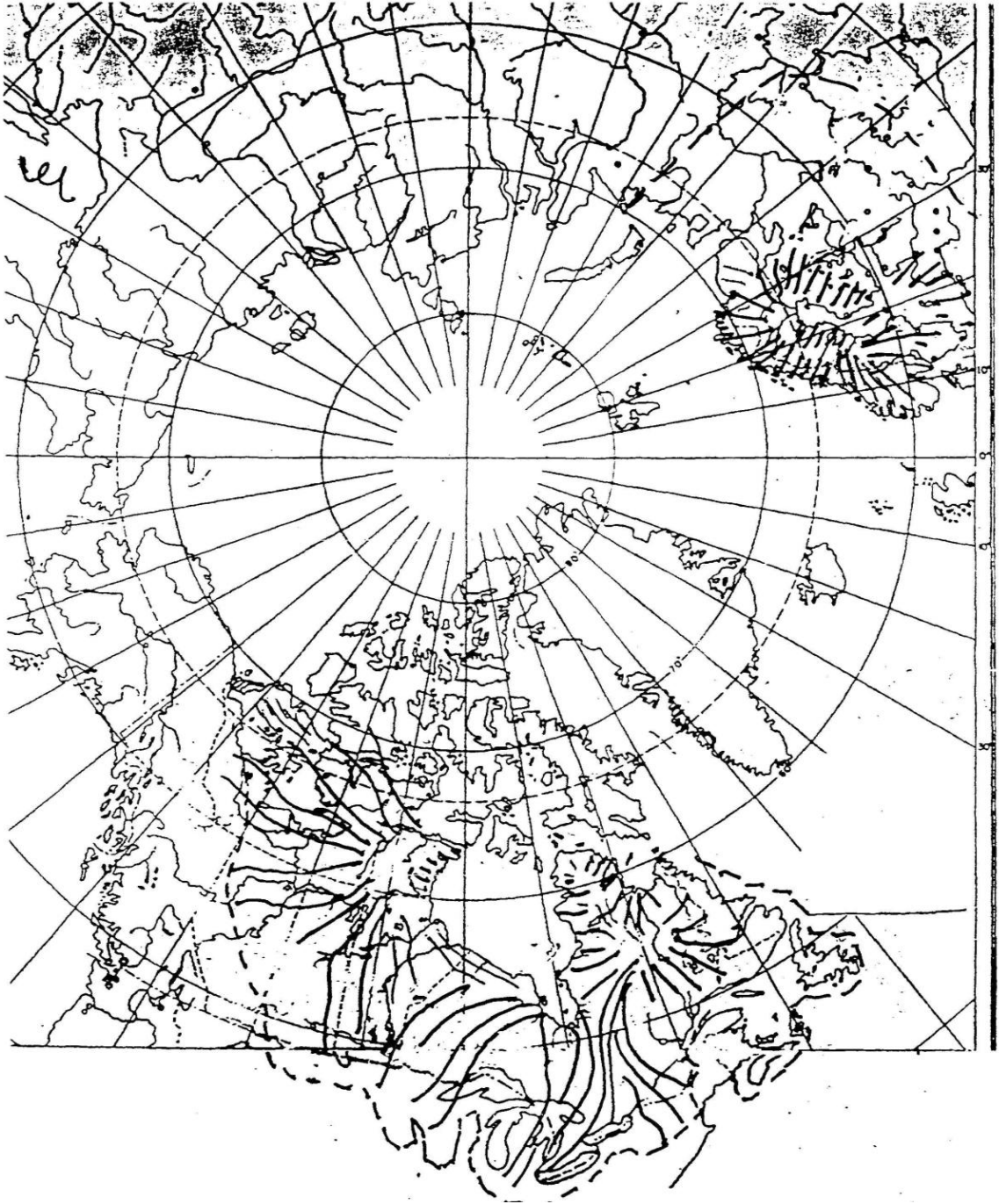

Generalized flow of drumlins and eskers in the final stages of the Pleistocene Ice Sheets of North America and Europe.

**Illustration 2**

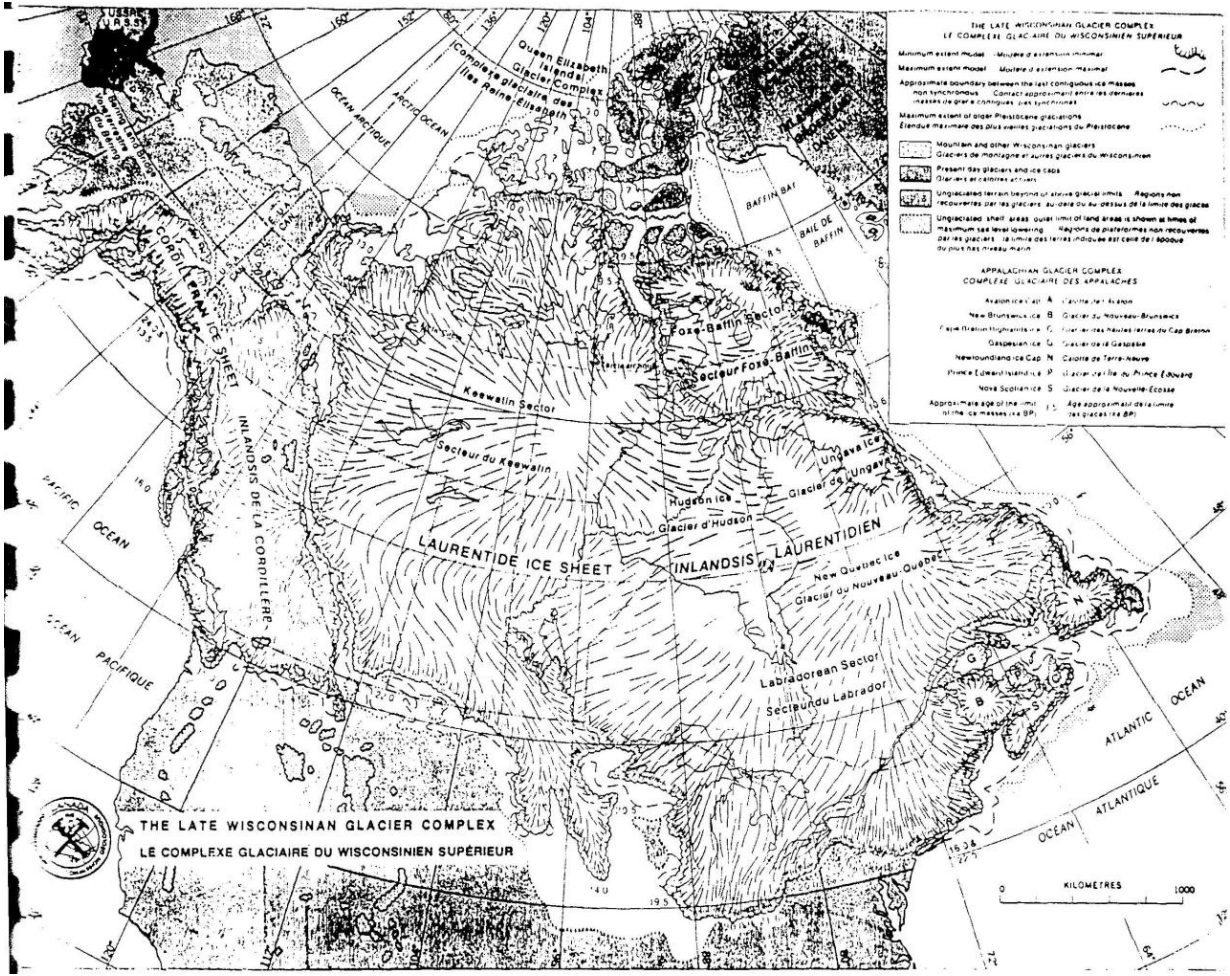

**Illustration 3**

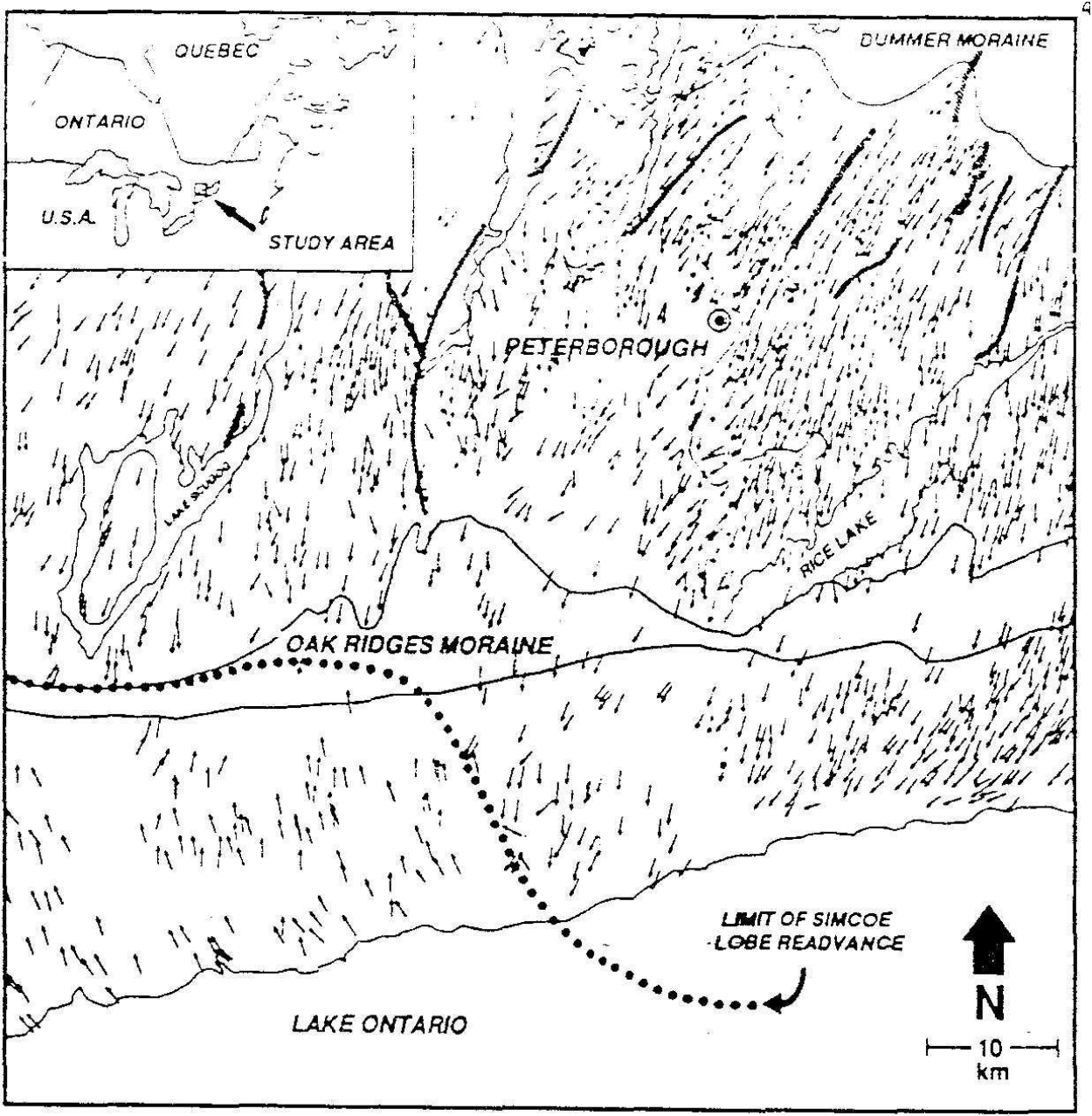

Distribution and orientation of drumlins (arrows) and eskers (lines) in Peterborough drumlin field. Drumlin formation occurred during readvance of part of Simcoe lobe (limit indicated by doted line), which overrode eastern half of Oak Ridges moraine.

**Illustration 4**

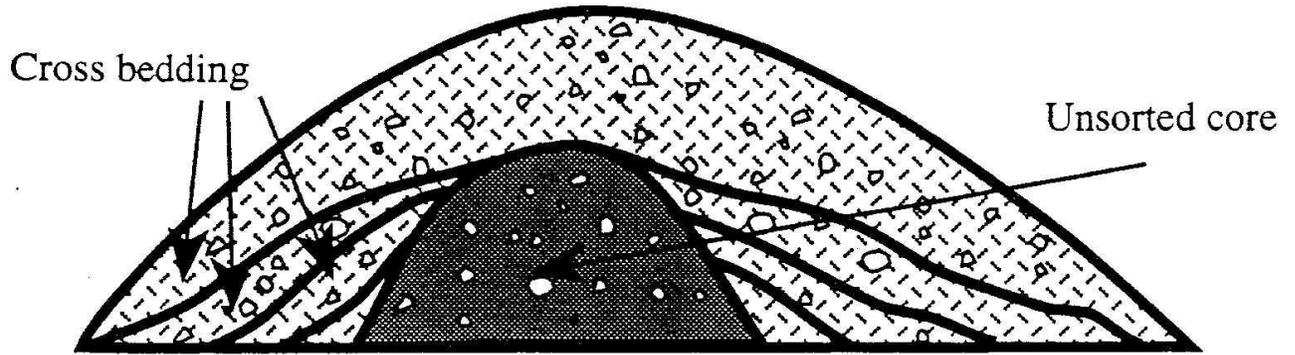

Cross bedding

Unsorted core

Cross section of a drumlin.

**Illustration 5**

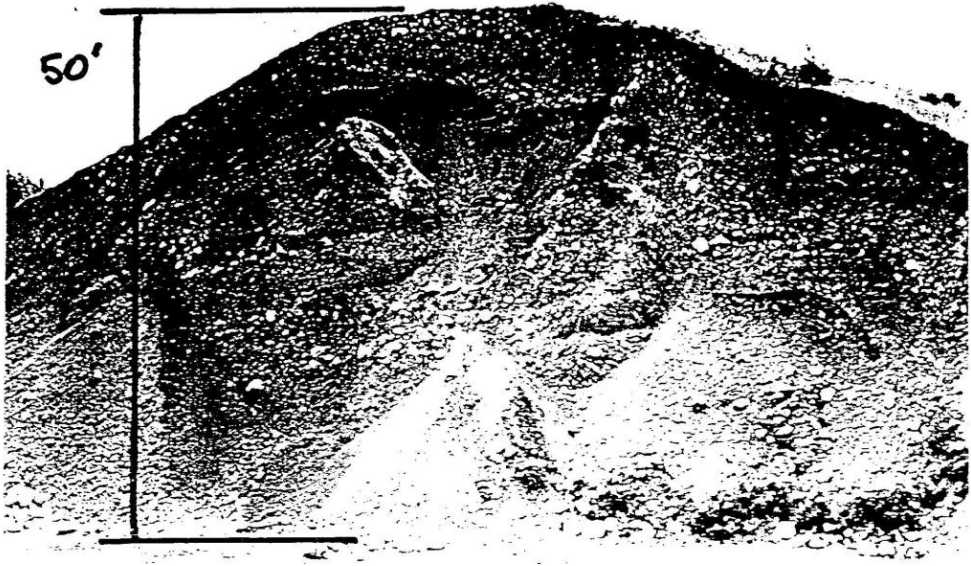

Pit face in esker, exposes the typical irregular bedding of gravel and sand. Notice cross bedding on the side.

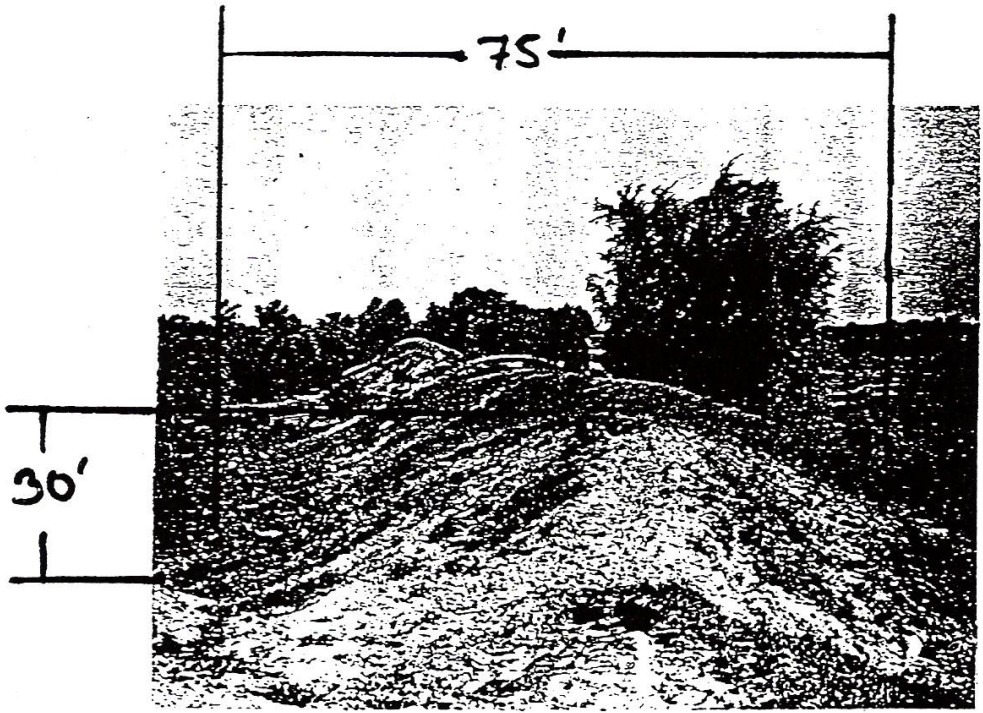

Esker near Atwood, south of Listowel. This small esker, and the gravel is unusually fine.

In North America they provide foresters with logging roads in swampy terrain. In Scandinavia, their closely-packed gravel and sand bedding is sound enough to provide roadways linking important towns 11. In the far north of Canada, wolves and caribou use eskers as raised pathways, occasionally meeting aboriginal hunters who hide behind nearby blinds. (12) (Illustration 6)

The ridges of sterile rocky rubble provide little return to the farmer. Aggregate producers, however, have a great affection for these sand and gravel mines. Moreover, as cities rise today, eskers disappear into the maws of concrete mixers and asphalt paving machines.

**Drumlins**

Drumlins and their associated structures are more numerous than eskers. A drumlin's interior is composed of the same unsorted rocky debris as an esker, but it is sometimes partially or fully cemented into rock. (13) Drumlins have a streamlined shape and display a prow-like feature that is not seen in eskers.

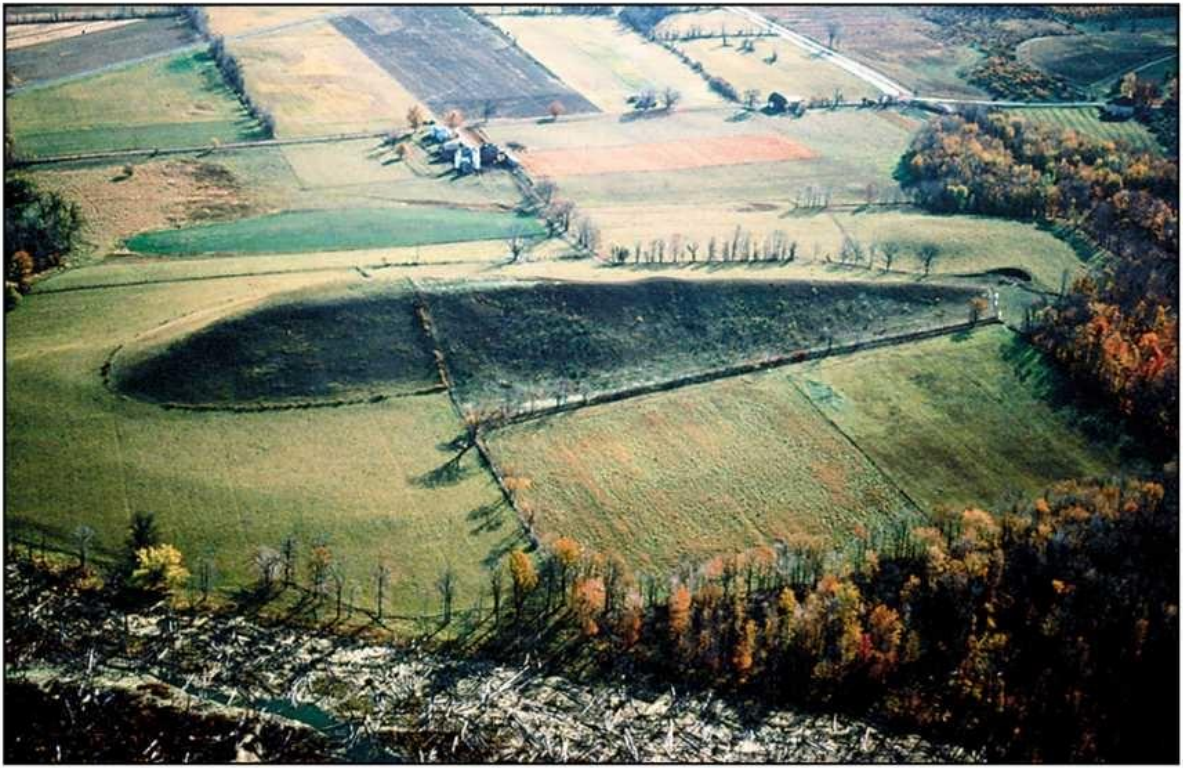

**The Mantle**

The distinctive ridge system that characterizes drumlins and eskers is frequently intimately associated with water-borne elements, usually fossil-rich clays and silts. These materials are sometimes absent from esker/drumlin swarms, but they are always considered a separate entity and often referred to as the mantle.

These mantling sediments may rise high enough to approach the ridge tops of a swam, forming platforms similar to continental and sub-oceanic plateau. These sediments usually favor the upslope side of the ridge complex forming annual layerings called varves, or run through ridge gaps or rise and flow over the ridge, covering the depressions in deposits. (Illustrations 4 and 5, above)

**Common Features**

Despite their similarities, eskers and drumlins are currently considered different geological phenomena and are hence explained separately. In fact, aside from slight morphological differences they share more similarities than differences. They are both elongated discontinuous ridges, mantled and often covered by the same type of water borne sediments. Under these mantles, drumlins and eskers show the same pattern of sterile, unsorted rocky debris. A very important distinction between ridge types is the degree of lithification i.e., cementing of the interior. The more numerous drumlins have more cemented rock, their aggregates concretised by calcium carbonate elements. (14) Other ridge types such as drumlinoids, drumlinoid ridges, flutes, craig and tail, roches moutonnees, whalebacks etc., show slight variations on the drumlin/esker model, but they are occasionally found eskers and drumlins in the same swarm. (Illustration 7)

Since all these ridge categories emerged from the same decaying ice sheet as members of a 'super-swarm', their shared birth, common habitat and close physical affinity suggests a similar conception.

We will argue in the following pages that esker/drumlin swarms were not generated by ice, but by fragmenting comets. These peculiar dual regimes of barren rocky ridge and adjacent fossil-rich ridge strata, will reveal an inner structure and topography consistent with a shallow landing of multiple comet tails. (Illustrations 8, 9, 10)

Before entering into a detailed argument for an extraterrestrial origin of an esker/drumlin swarm, let us first consider current explanations of its origins.

**Illustration 6**

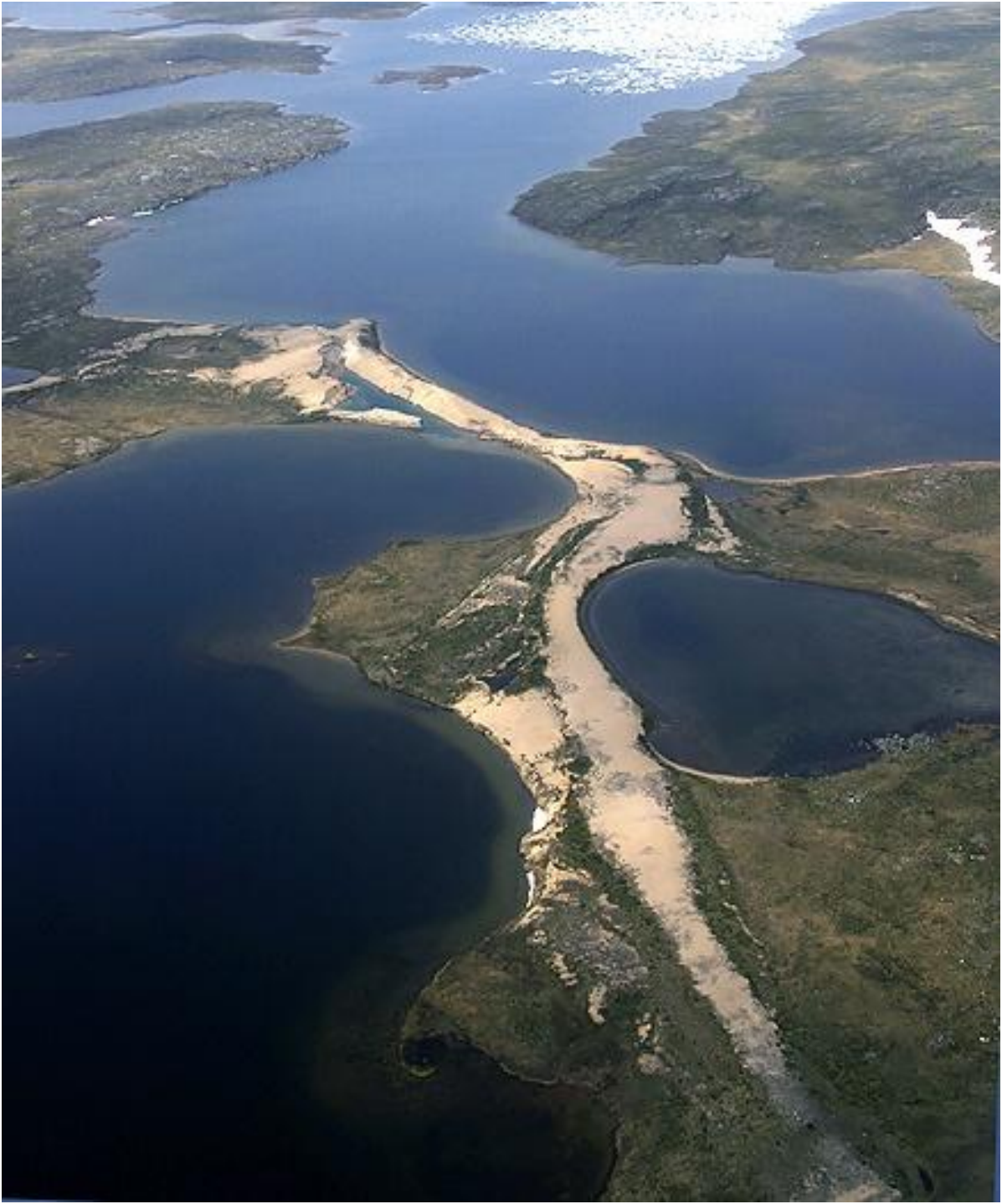

Photograph of esker in the Barren Lands, Nunavut.

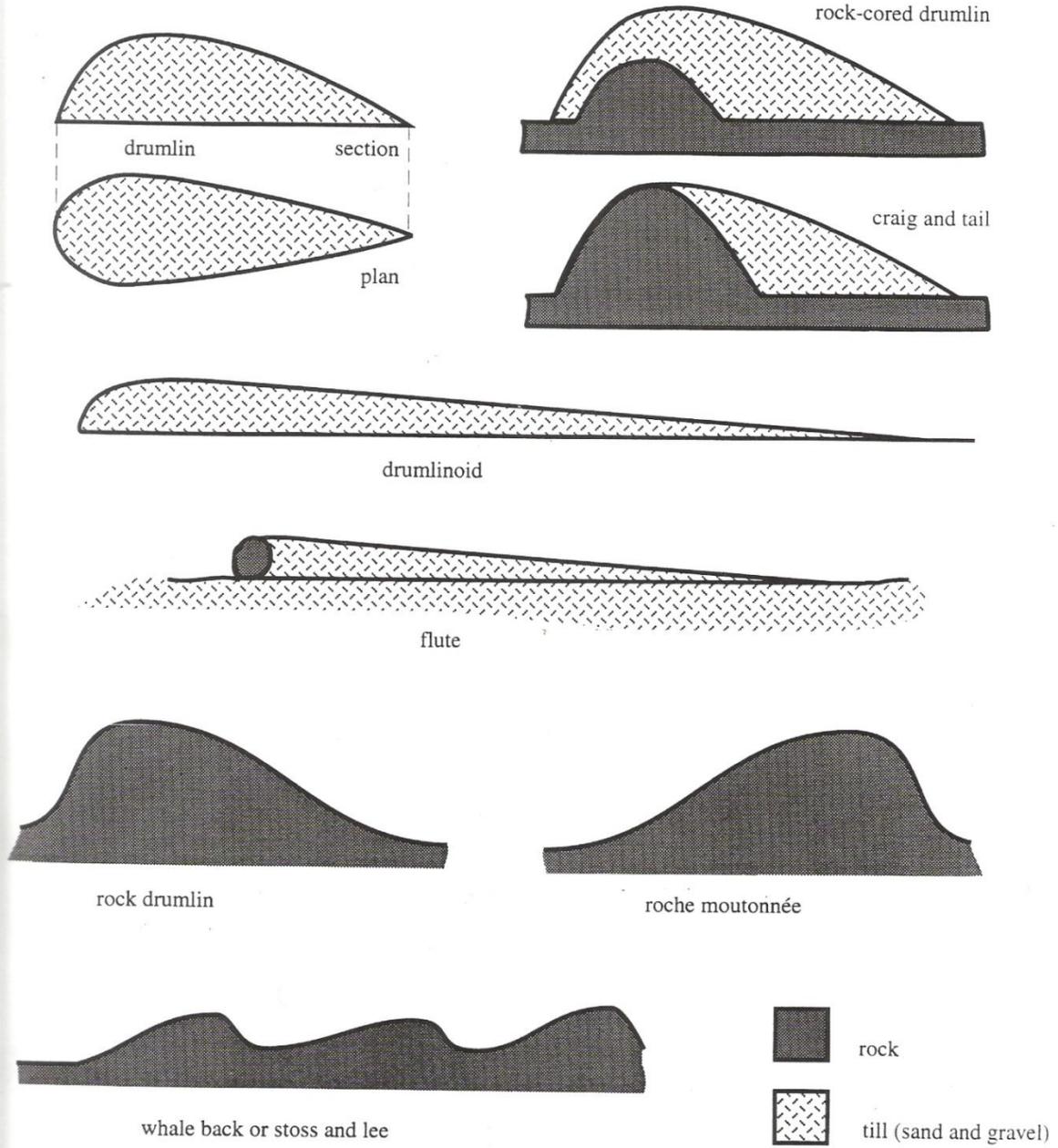

**Illustration 7**

## Glacially Streamlined Features

Direction of ice movement →

rock-cored drumlin

drumlin   section

plan

craig and tail

drumlinoid

flute

rock drumlin

roche moutonnée

whale back or stoss and lee

rock

till (sand and gravel)

## Illustration 8

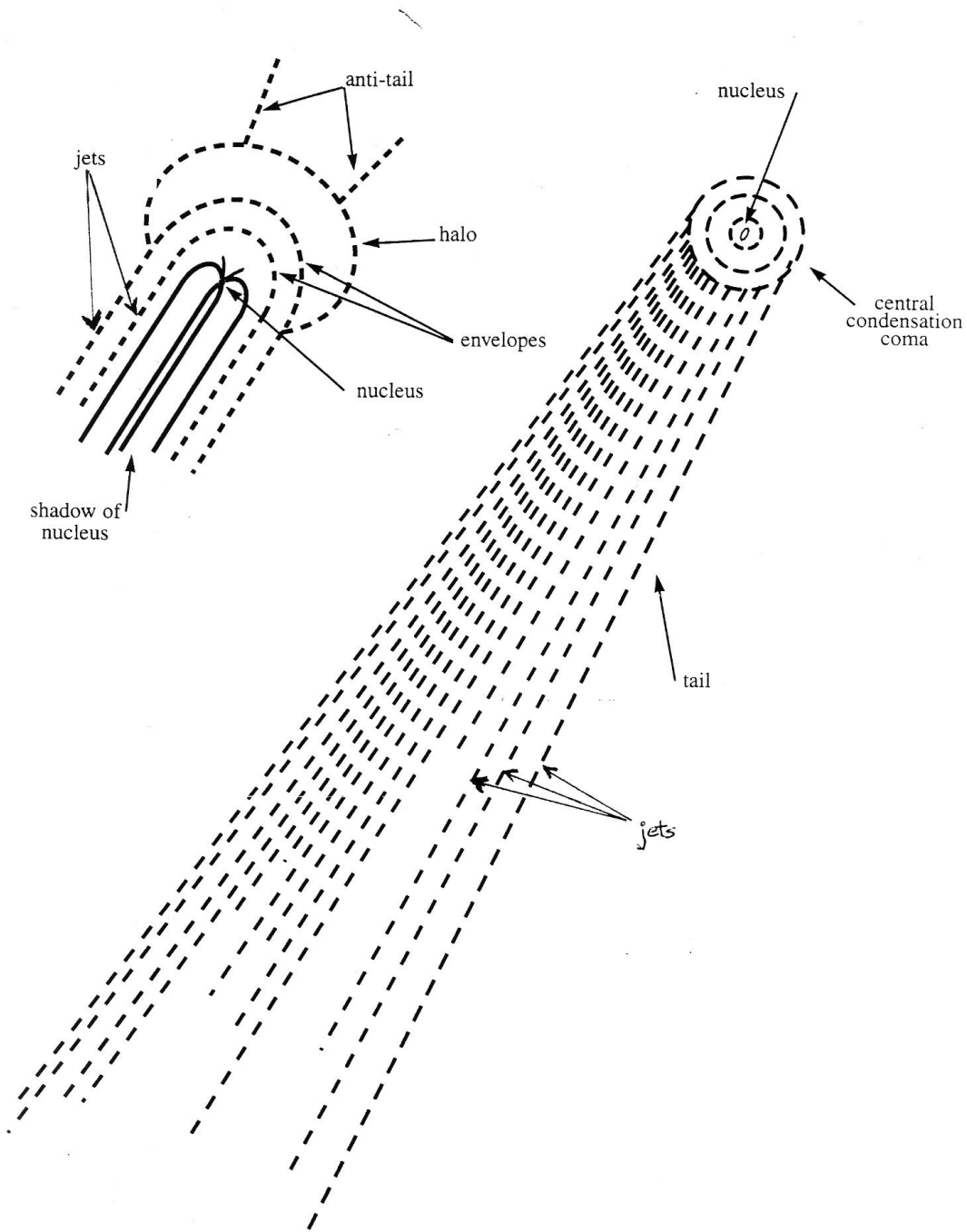

Diagrammatic representation of the parts of comets.

**Illustration 9**

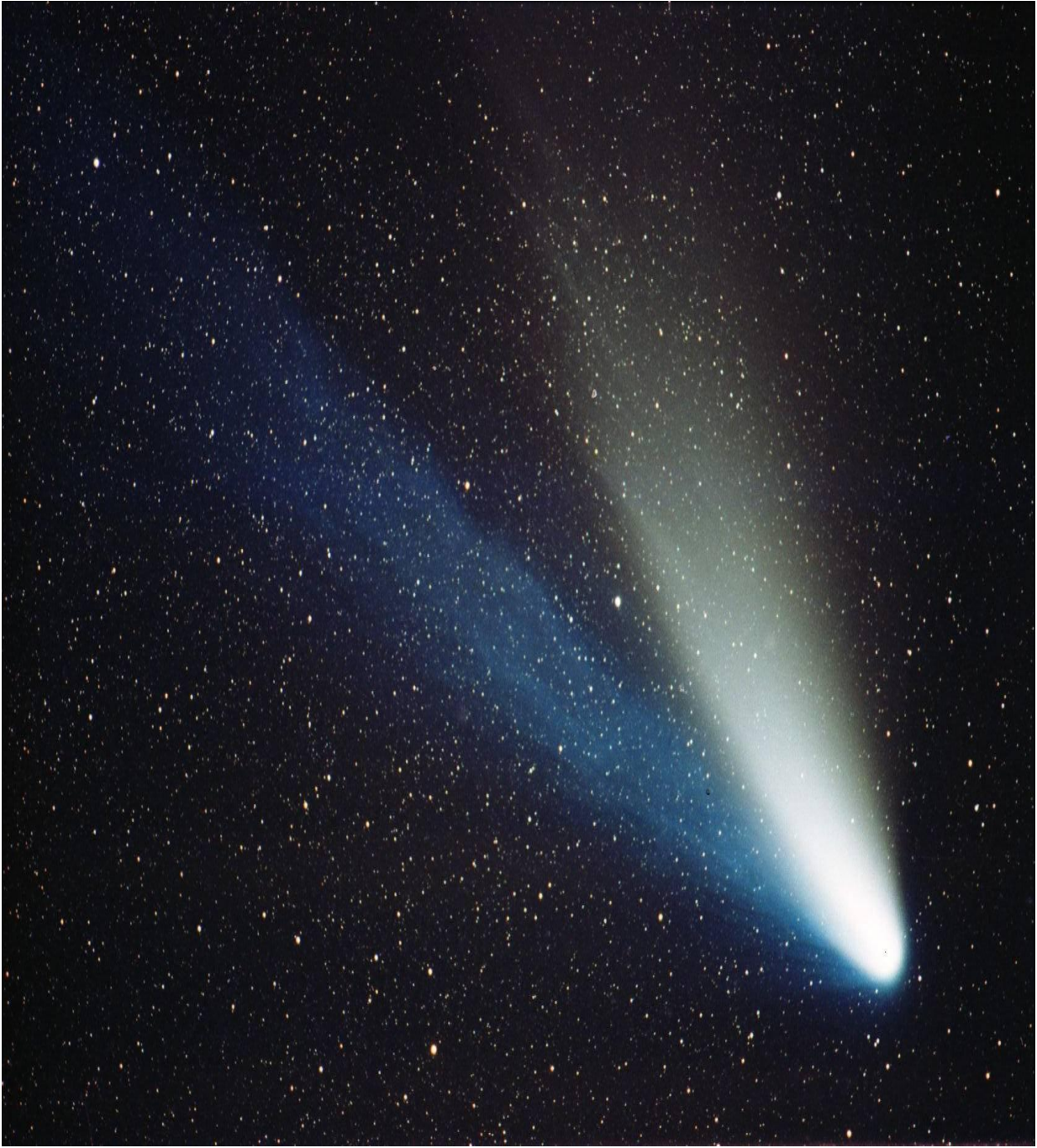

**Comet Hale Bopp.**

## Illustration 10

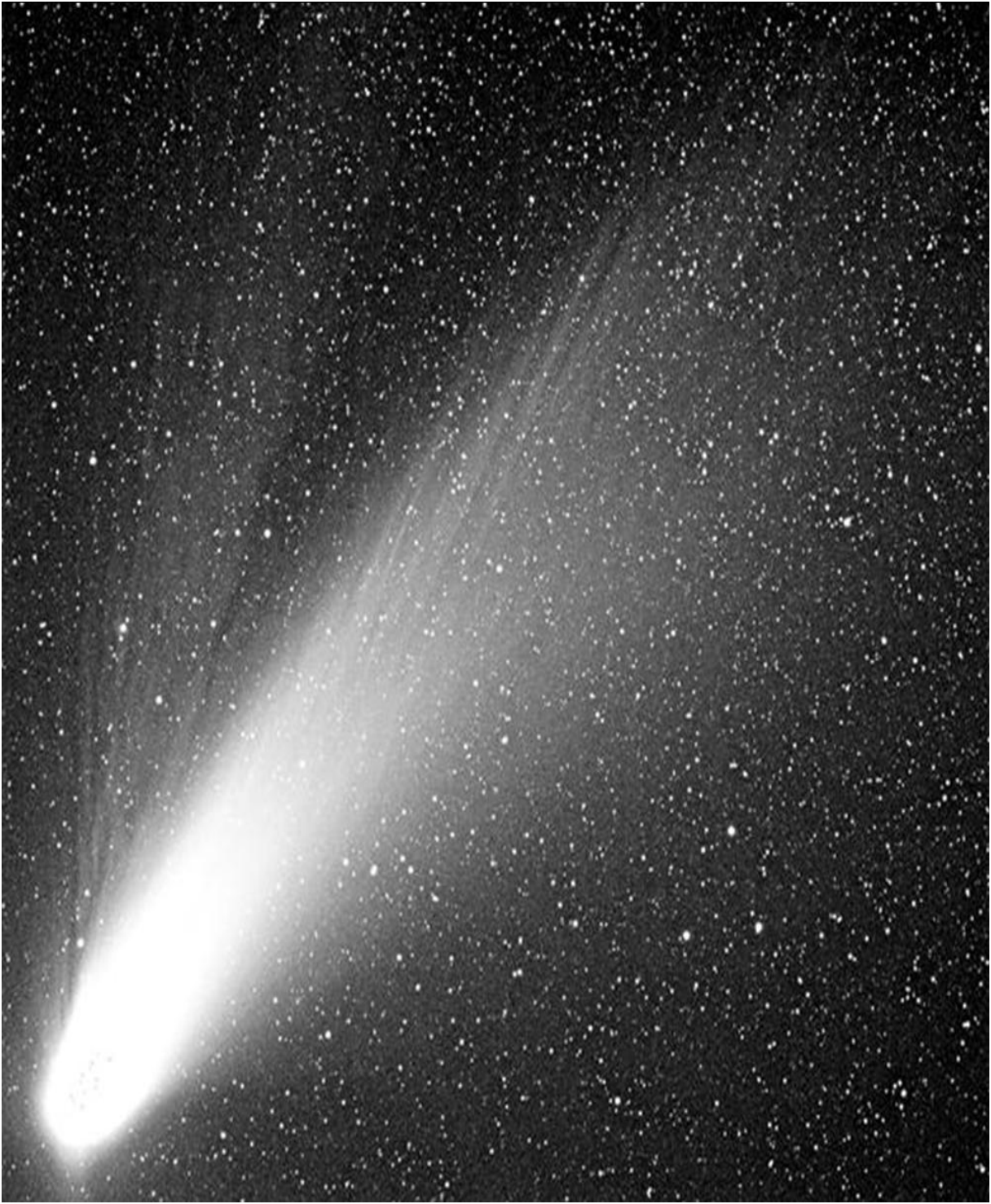

**Comet Hale Bopp.**

# EARTH-BOUND SOLUTIONS

## Hummel's Explanation

D. Hummel, a nineteenth century geologist, made one of the first attempts to understand the esker ridge phenomenon. "Most eskers" he wrote, "are the deposits of glacial streams confined by walls of ice and left as ridges when the ice disappeared." (15)

Associating eskers with ancient melting ice sheets explains their excellent state of preservation. Any densely packed ridge of gravel and sand lying on top or within a melting ice sheet, should preserve its shape, as the structure is uneventfully lowered onto the solid ground. There, the exposed complex would ultimately lie, unaffected by gentler topographies, draped over hills and crossing watersheds. There seems little question that eskers and drumlins are the product of the retreat of the ice sheet that last covered a large part of North America and Europe. Problems arise when geologists attempt to explain how the debris originated, was transported, and laid down.

## Conventional Arteries

Billions of cubic meters of rocky debris are moved downstream to the sea by the world's rivers. An equally complex mixture of debris moves downslope in Alpine glaciers. Both systems share some important similarities. Their tributaries join together to form a main artery which bears its loads further downslope in a visually apparent progression. Neither system of channels, however, leaves behind sedimentary debris resembling the characteristic pattern of an esker swarm.

In rivers and glaciers, individual particles are mechanically worn down by transportation. Moreover, all transported debris should have a fossil and organic content similar to the topsoil and bedrock through which it travelled. All these attributes of fluvial and glacial transportation are conspicuously absent from esker and drumlin interiors. In short, the architecture and contents of eskers do not suggest any connection with the mechanics and processes of existing valley glaciers or river systems.

**The Ice Age**

The greatest difficulty in finding an appropriate model for esker/drumlin swarms lies in the simple fact that they are not currently forming. (16)

Hummel, at the time he first examined eskers fortunately had a newly established paradigm to turn to — continental glaciations. Under their mantle of sediment, esker and drumlin ridges, as drift artifacts, contained till, or boulder clay (as it was occasionally named in the nineteenth century), and since till was thought to have been ground out by continental glaciers, it could only be a by-product of the ice age.

It was these vast expanses of ice that the renowned nineteenth century naturalist, Louis Agassiz, embraced in his search for a vehicle for ending earth ages. By the time Agassiz began his work in Neuchatel, in the late 1830s, local naturalists had already begun speculating on the origin of 'erratics'. These large broken blocks and similar trains of smaller debris were sometimes found hundreds of kilometers from their suspected origins.

Competition to explain their mode of transportation developed between two factions; the deluvialists and the glacialists. The deluvialists, headed by Charles Lyell, believed that great floods and large icebergs17 had transported them. A new group, the glacialists, headed by Jean Charpentier, a naturalist who had taken many field trips with goat and mineral hunters into the glaciers of the Alps and Jura mountains, developed the idea that the ability of glaciers to gather, convey and grind rocks could be applied to areas outside existing Alpine valleys. (18)

Originally skeptical of the power of glaciers to convey erratics, Agassiz later became so enthusiastic a supporter that he is credited with the establishment of the ice age as both scientific canon and popular belief. He went so far as to claim that ice totally covered both hemispheres, relentlessly bringing each succeeding earth age to its end. Though main-stream geological thought never accepted Agassiz's catastrophic ideas or the earth's total immersion in ice, by the time Hummel had examined his first esker in 1874, the continental ice sheets were the accepted generator of most of the surficial debris, still known as drift, found on northern continents. Geologists succeeding Hummel and Agassiz proposed a variety of glacial 'techniques' to grind, collect, form, carve, transport, elevate and lay down the complex formations found in an esker/drumlin swarm.

In many ways, a valley glacier is an ideal engine for generating both till and sediment. Its great rivers of ice are capable of carrying deposits of debris on their surface and sides and of using the smaller and harder rocks to scour and polish the channel walls as they flow into the Alpine valleys. It was therefore logical to attempt to push glaciers beyond their confining valleys and extend their new reach to explain an esker/drumlin swarm.

Even assuming that the mountains of pre-history were higher than at present, we would still expect to find their broken rubble confined to traditional pathways. On the contrary, ridge swarms, as mentioned above, tend to ignore these pathways. The major divides connected with most drifts do not radiate from mountain ridges and there are no divides centered on or near the Rockies, North America's greatest ridges. The European divide is actually centered below the water in the Gulf of Bothnia.

Moreover the putative products of valley glaciers are found scattered all over the summits, and sides of mountains, far from the beds of alpine valleys that supposedly conveyed them.

Agassiz proposed a radical solution to this dilemma. He posited an already ice covered earth suddenly punctured from below by great mountain ranges. Great mountain chains such as the Alps broke through the thick ice cover and scattered rocky debris upward and outward, linking mountain and distant plain with the requisite trains of debris.

Agassiz's big bang, was ultimately rejected and forgotten, but apparently left the glacial origin of till intact. To this day geologists when examining great esker/drumlin swarms do not trace them directly to the so-called great ice divides from which they supposedly originated. This is probably due to a continuing ambivalence with regard to the ability of ice sheets to convey tills any significant distance. In a recent text Jurgen Ehlers acknowledges this problem, admitting the short haul character of an ice sheet, but prudently adds the observation of trails of debris hundreds of kilometers long which appear to possess a connective character.

## The Mid-Stream Argument

Hummel's conclusion that eskers were glacial stream deposits may be plausible. After all, the ridges are often covered by extensive alluvial deposits. Yet the esker and drumlin ridges lying under this water-borne debris could not have been laid down by alluvial processes, and this anomaly lies at the heart of the esker enigma.

## Inner Core and Outer Mantle

Eskers and drumlins are accompanied by stratified beds consisting mainly of clay and silt that sometimes completely bury the ridge structures. But these truly fluvial deposits were obviously formed in a sequence and by a method different from its companion ridges. (Illustration 4, above)

The first to observe the distinction between the core and its mantle was Kropotkin. On examining the contents of an asoar (esker), excavated at Upsala, Sweden, in 1897 he wrote, "A strict distinction must be made between the… core… of an asoar (esker) and its mantle. They are of distinct origin. The latter is always due to action of water (rivers, lakes, or the sea), while the core, whenever access could be found to it, was… always... found to consist of unwashed and unstratified till, and never of fluviatile deposits. This core is often buried under a thick sheet of water-deposits, and occasionally it lies even beneath the level of the surrounding plains. It must have the same origin as drumlins, ... (which) always accompany asoars (eskers)." (19) This last comment of Kropotkin is included, for it is one of the rare statements concerning the obvious genetic relationship of drumlins to eskers, an observation that is almost impossible to find today. (20)

## Grinders and Streamers

The explanation of an esker/drumlin swarm generally depends on the regime, inner ridge or outer mantle, chosen for study. If it is the inner core or ridge element, a geologist tends to favour moving ice. This judgment reflects the barren and unstratified condition of the interior. If it is the adjacent stratified mantle, it is water.

One can understand the enthusiasm of those choosing the fluvial model, for frequently water-borne sediment is all that can be seen of an esker/drumlin swarm, particularly where the ridges collect enough debris to form wide raised belts or mini-plateaux. We shall call this hydraulically inclined faction the 'Streamers'. There is, however, a contingent of geologists who still believe both eskers and drumlins to be essentially formed by the action of retreating and advancing ice sheets. They point to the high content of broken rubble and sharp-edged sand that can only be the result of glacial grinding. (21) This group, whom we will call the 'Grinders', still uses the existence of erratics to establish the direction of continental ice sheet movement. (22) Erratics, large rocks or trails of loose rocks resembling those in more remote bedrock, have been used by both Streamers and Grinders to support their genesis theories. (23) The glacial grinding hypothesis mandates the gouging, mixing and movement of enormous amounts of rubble over great distances up and over significant elevations, a physical demand which has been seriously challenged for close to a century. (24)

The most sophisticated theories of esker formation employ a two-stage process to explain the inner core and outer mantle found in eskers. In this analysis it is presumed that the inner core is gouged out in the advance of a glacier and the water-borne debris is deposited in the retreat, each phase employing the same conduit. In some approaches, to explain the resulting 'braiding', an enclosed conduit is abandoned for open channels originally on the ice sheet's surface, apparently using existing ridge structures as a path. (25)

**Smooth among the Rough**

To add to the difficulty of using current accepted glacial models to explain esker/drumlin formation, the aggregate composition of ridge interiors contradicts the ideas of both Grinders and Streamers alike. Mixed in with the course gravels and sharp-edged sands, are larger rocky elements – cobbles and boulders — which show extensive erosion. These rocks, which are rounded and described as water-worn, are found in the ridge and not the mantling sedimentary regime. The co-existence of both smooth and sharp debris is inconsistent with a glacial or a fluvial model. As Kropotkin noted, the typical signs of water transportation, such as stratification and size sorting are not in the basic ridge structures. They are seen only in the mantling sediments and aside from the orientation of larger stones to the ridge direction, the ridge's contents are unstratified and unsorted.

The Grinders, who point to the rough texture of sand and gravel as proof of their hypotheses, must first find a way to grind, collect, transport and then levitate this debris into some kind of watery medium which must selectively abrade only the larger rocks and then lay the collection of core rocks down… unsorted.

Even if we posit that ice or water transported and formed eskers and drumlins, we must accept that the mechanics of excavating and elevating the rocky debris over local elevations and then transporting these materials hundreds of kilometers, are not only theoretically problematic but at present unknown in nature. This latter condition is prohibited by the first law of geology: that no process may be invoked to explain our past that is not currently observed at work (26).

## Formidable Problems

Another dominating dynamic and structural anomaly exhibited by an esker/drumlin swarm is its core's sterility. Its 'barren' condition was reported by early investigators. (27) This problem was not confined eskers and drumlins. How could any transportation system, gathering materials from hundreds of square miles, in a hinterland loaded with fossil-bearing rocks and other organic detritus, not leave its traces in the core interior?

Just as vexing is the partial concretization of some of the ridges found in swarms. The conversion of rocky ruccle to sedimentary rock is still believed, by many geologists, to have been accomplished through a combination of heat and pressure. (28) This 'metamorphic' process was believed to produce complete mountain ranges through cycles of vertical submersion, and emersion. Recently challenged by the discovery that mountain ranges are superficial structures lying directly on older strata, geologists are still turning to other pressure sources such as plate tectonics to explain the production of sedimentary rocks. Esker/drumlin swarms, however, are very recent, and having never been pushed into the bowels of the earth, or squeezed by moving plates. Their relatively rapid conversion into rock cannot be explained by conventional means.

**Ridge Lithification: Binding the Earth**

Much of the earth's exposed rock originated as loose sediment. Efforts to explain how these loose aggregates of sand and gravel reconstituted themselves into solid crust have been only partially successful. The earth's softer rocks, its shales, or mudrock, may be compressed into easily broken laminates by means of pressure. The majority of rock, the sandstones and limestones, are essentially cemented sand and gravel, a chemical and physical composition that is indistinguishable from concrete. There are aggregates such as clays and volcanic debris that can be heated or precipitated into rock. Volcanic activity is capable of producing cements in its oxygen free vents, but this material is not found in eskers. Valley glaciers create a powdery material called rock flower, as a by-product of their grinding activity. This material, though not found in eskers, might be presumed to be the cementing agent for a swarm. Rock flour, however, is not produced in what is called a reduction (low oxygen) environment, and is hence not capable of cementing.

The rock formations found in an esker/drumlin swarm are not volcanic in origin. Nor could the enormous quantities of cements that form the ridge of esker/drumlin formations have been produced by the repeated gouging and grinding that created till.

The enigmas of the esker/drumlin's sterility and its tendency to turn to rock quickly (in geological terms), while representing intransigent problems for an earth-bound solution, lend themselves ideally to an extra-terrestrial explanation.

The limestone-rich debris ejected from disintegrating comets, projects its dust into an oxygen-reduced environment. This idea was touched on by John Penniston, an astronomer, who wrote on northern China's loess formations. (29) Loess is a silt or silty conglomerate found covering the surfaces of most hemispheres. In large areas of northern China, it is found in the form of a low density conglomerate. Spike-like tubules of calcium carbonate formed by water trickling through the loosely packed material create a lightweight material which is equally useful for carving out dwellings and fertilizing fields. Penniston noted that the large amounts of calcium carbonate used to hold the conglomerate together could only have come from calcium oxide. This chemical popularly known as slaked lime was used as the cementing agent in concrete production. The astronomer noted that calcium oxide, though rarely found in natural form on earth, was a common component of meteorites. We contend that the source of the calcium carbonate that holds loess together is the same material that binds esker/drumlin ridges together. (Illustration 11)

**Core Content**

Present theories of esker formation assume that a large portion of the rocks, gravel, and sand in the ridge formations were derived from materials gouged out, ground up and gathered from bedrock and surface debris. This coarse and broken rock is presumed to be till. (30)

Its nearby abundance, in both scattered and bedrock formations, gives support to this view. The Streamers however, note that eskers and drumlins lie directly over undisturbed surfaces, leaving soft local bedrock and sediments unmarked by their passage. (31) These investigators have felt justified in favouring Hummel's original fluvial hypothesis. (32) In defense of the Grinders, it must be noted that even Streamers believe that virtually all the contents of a swarm were originally manufactured by ice. Since glacial grinding is one of the few dynamic processes claimed for swarm genesis that is currently observable, we will turn to it next.

**Illustration 11**

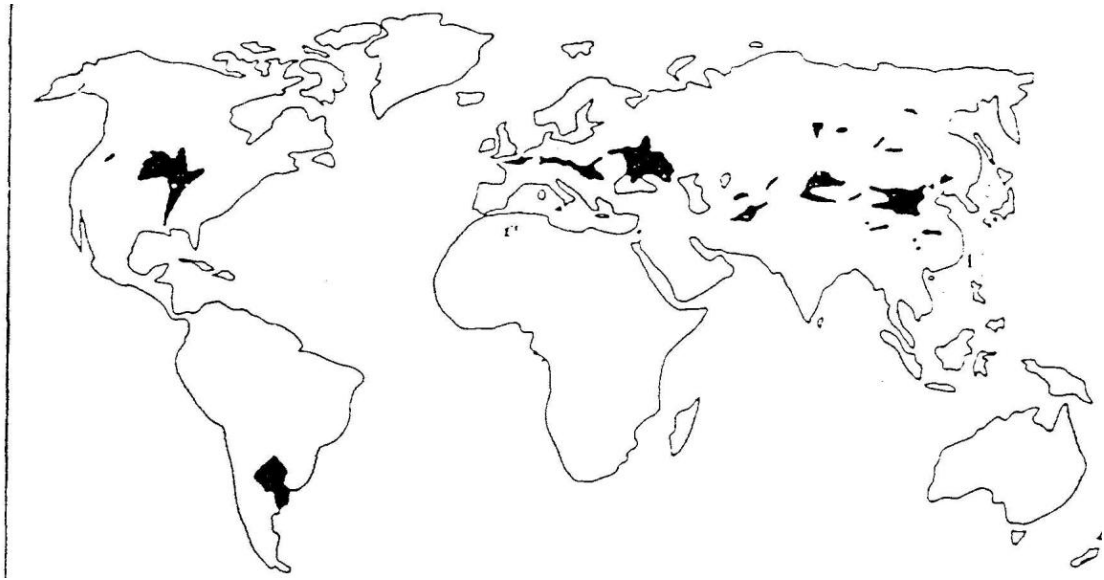

**Map showing global loess deposits.**

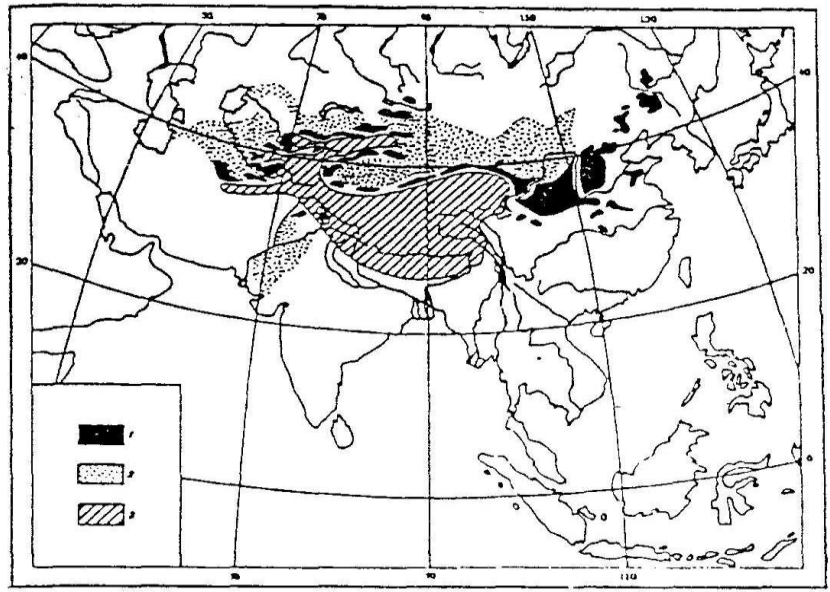

Sketch map of the distribution of Loess in central Asia:
1. Loess cover
2. Sandy deserts
3. Mountains up to a height of more than 3000 m (data for China from Liu Tung-sheng "Loess and the Environment" Beijing 1985)

**Alekseev-Dodonov map showing loess in the uplifted region of Central Asia.**

**Sourcing Eskers and Drumlins**

Geologists use existing valley glaciers as an example of the techniques ancient ice sheets used in generating most of the core materials found in an esker. Rocky debris, falling from mountain peaks, gathers on tops, in the sides and bottoms of advancing valley glaciers. This rubble carried by the heavy moving ice, abrades and scours its tributaries and main channels. The dynamic behavior of valley glaciers is shown in Illustrations 12, 13, 14. But neither the swarming features of an esker/drumlin swarm, nor its architecture is found within glacial valleys. Only an occasional stubby, deformed ephemeral, orphan ridge is generated in Alpine sites. Moreover, the more numerous cemented companions of eskers — flutes, craig and tails, drumlinoid ridges etc., -- are neither found nor generated by present glaciers. The life of these Alpine glacier mini-eskers is short for the retreating glacier that temporarily exposes a mini-esker ridge in a warm season, will wipe it out in the next advance. The mud flowing through fractures in the leading edge of glaciers leaves inclusions in dirty ice, which melt in warm weather, leaving the formation distorted and slumped. These 'eskers' are related to the debris formed by valley glaciers called terminal and lateral moraines and kame terraces that may be seen in any valley glacier. The same forces which deposited mixed ice-filled debris on the sides and termini of glaciers will occasionally gather that same material in the lowest point in the valley floor and temporarily expose it during a period of retreat.

It is generally agreed that these valley glacier artifacts are not related to the swarming feature we are examining in this paper. It is necessary to underscore this point for many textbooks state that eskers have been found in mountain valleys. In fact one of the few places where esker/drumlin swarms are not forming is within past and present glacial valleys. The examples cited as eskers and drumlins are exposed by retreating glaciers on valley floors, and may be examples of vestigial ridges laid down and then covered by an advancing glacier. (34)

**Influence of Topography**

Although eskers and drumlins appear to ignore topography, one leading glaciologist has taken an ambiguous position on the issue. His view casts an important light on the problem of ice sheet dynamics. Richard Flint has stated that "where conspicuous valleys are present some eskers follow the valley floors, but not where valleys tend to diverge greatly from the direction of the former glacial flow." (35)   At first these observations appear contradictory, since Flint himself admitted that esker/drumlin swarms routinely make serious ascents over the hills and streams of Europe and North America. (36)

**Illustration 12**

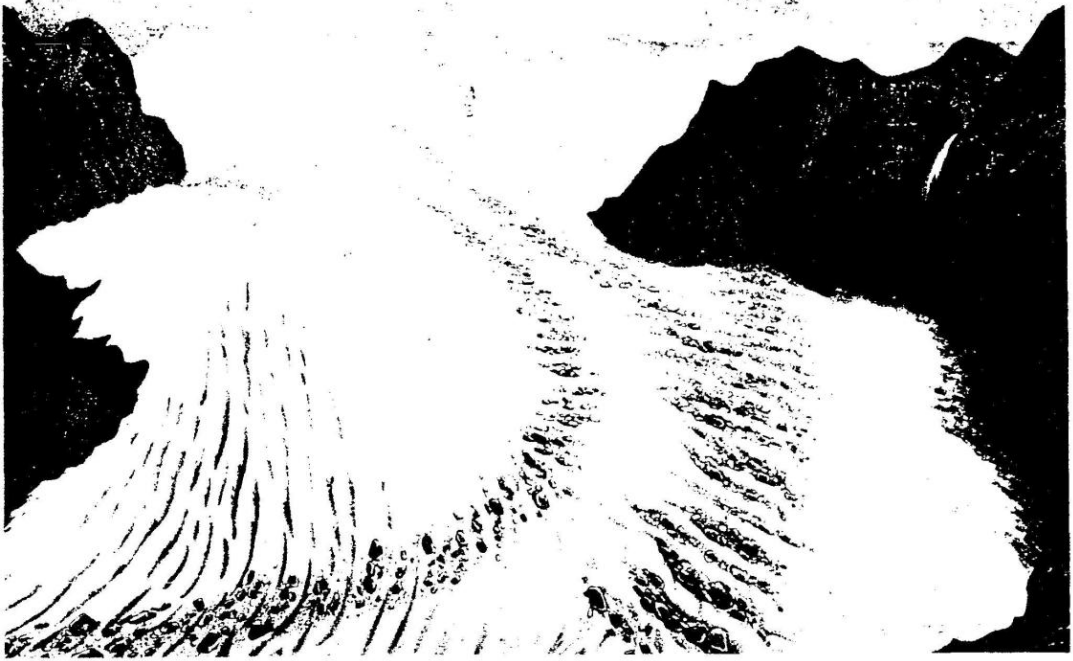

**Figure 1. Lower Part of Zermatt Glacier with medial moraines (Reproduction of Plate 5 in Agassiz's *Etudes sur Glaciers*, 1840).**

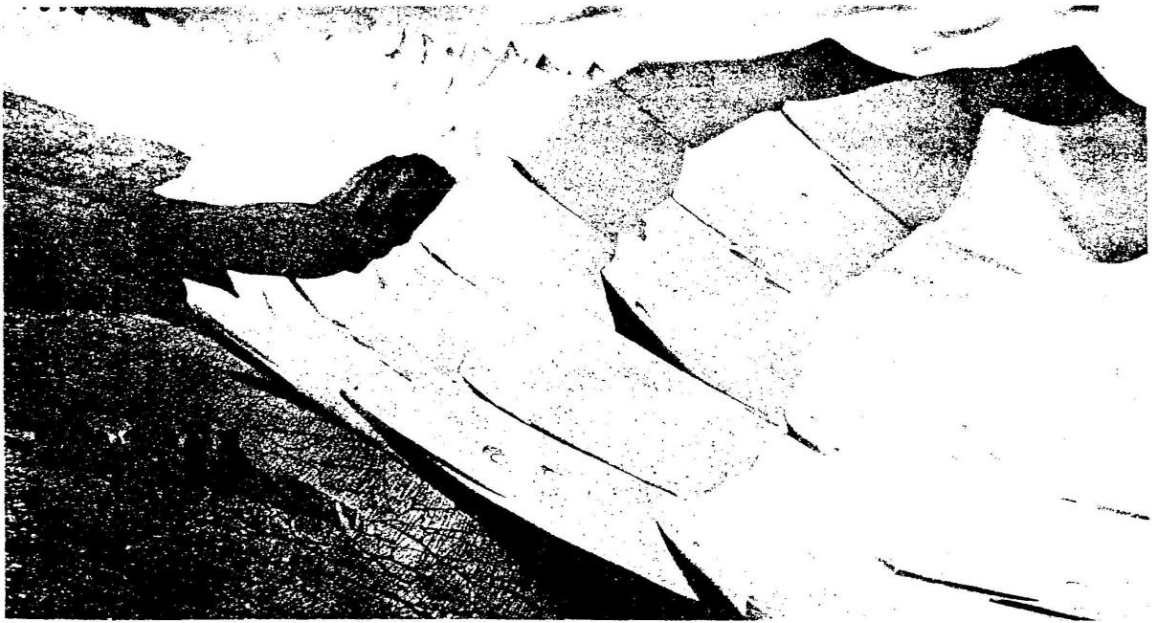

**Figure 2. Side View of frontal part of Zermatt Glacier on polished and striated bedrock (reproduction of plate 7 in Agassiz's *Etudes sur Glaciers*, 1840)**

Illustration 13

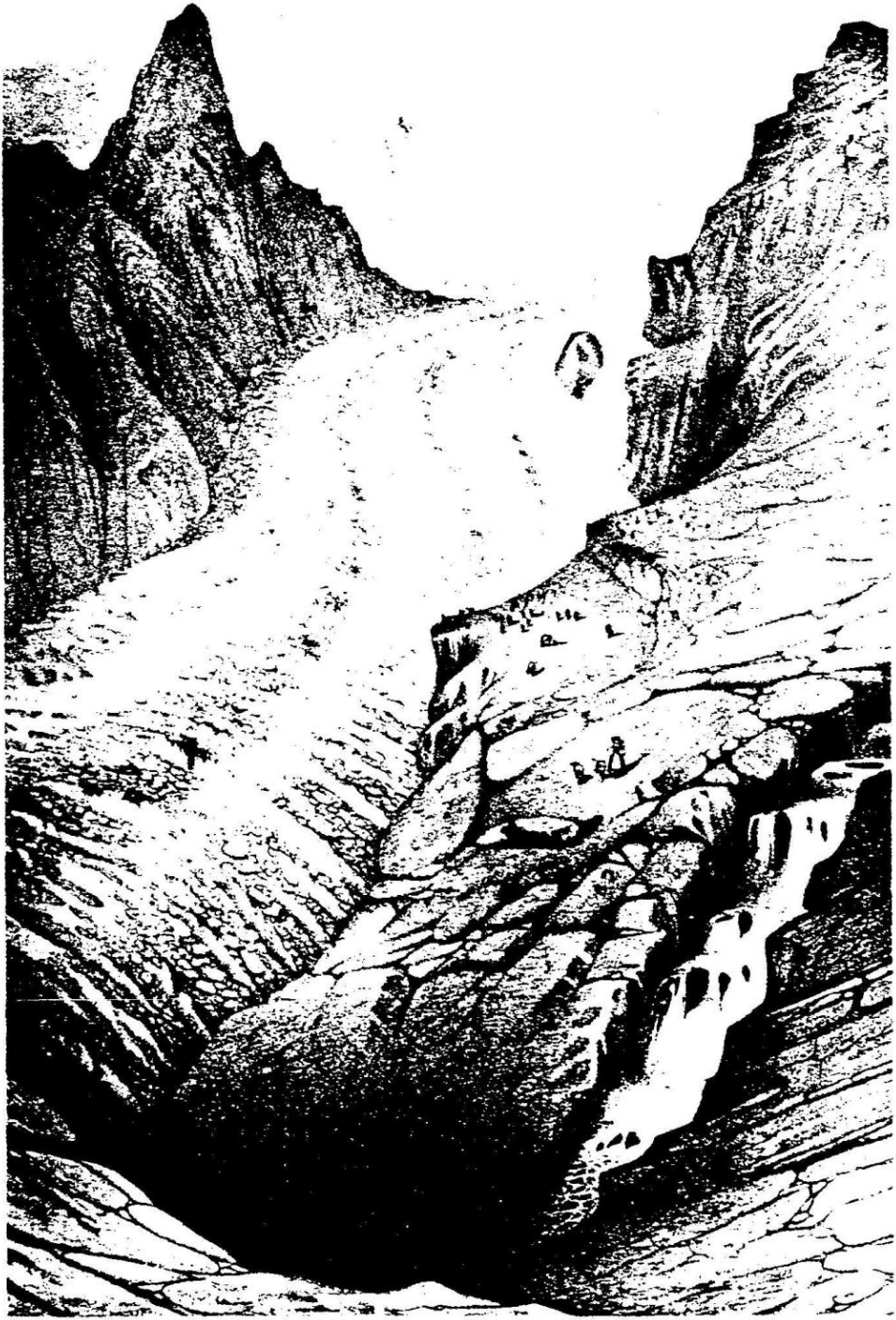

**Middle Part of Zermatt Glacier in its U-shaped valley and with lateral and median moraines (reproduction of Plate 4 in Agassiz's *Etudes sur Glaciers*, 1840)**

**Illustration 14**

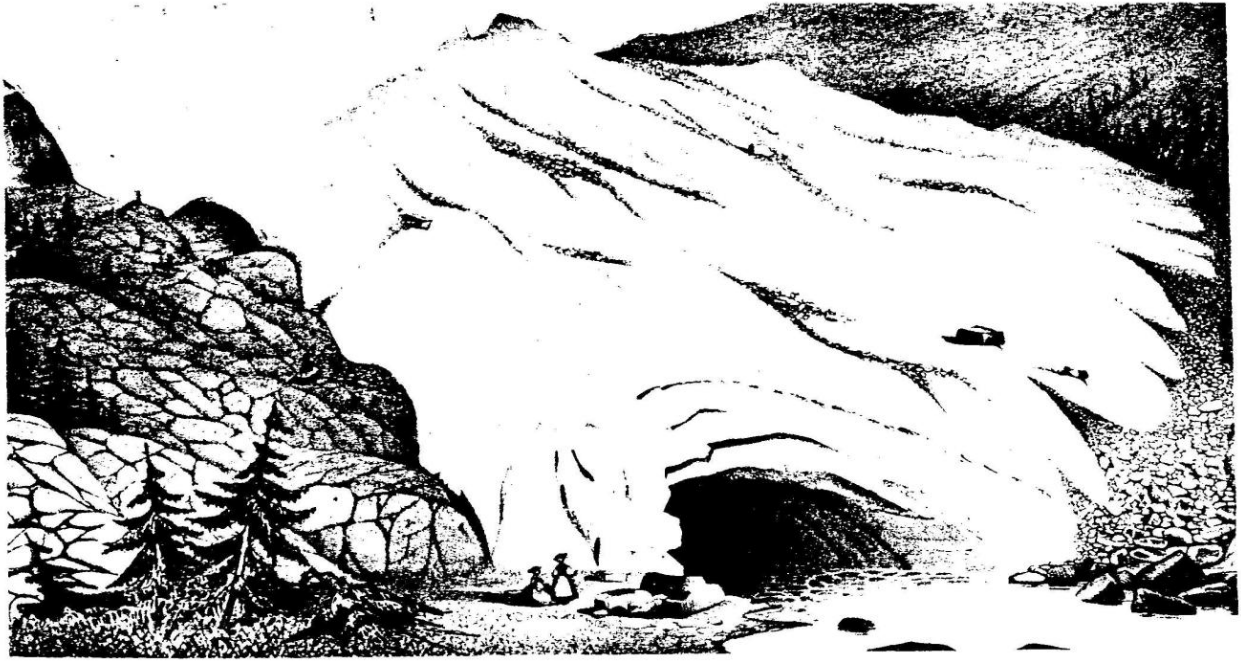

**Figure 1. Front of Zermatt Glacier with "roches moutonneés" on the left, ice cave with subglacial stream in the middle and lateral moraine on the right (reproduction of Plate 6 in Agassiz's *Etudes sur Glaciers*, 1840).**

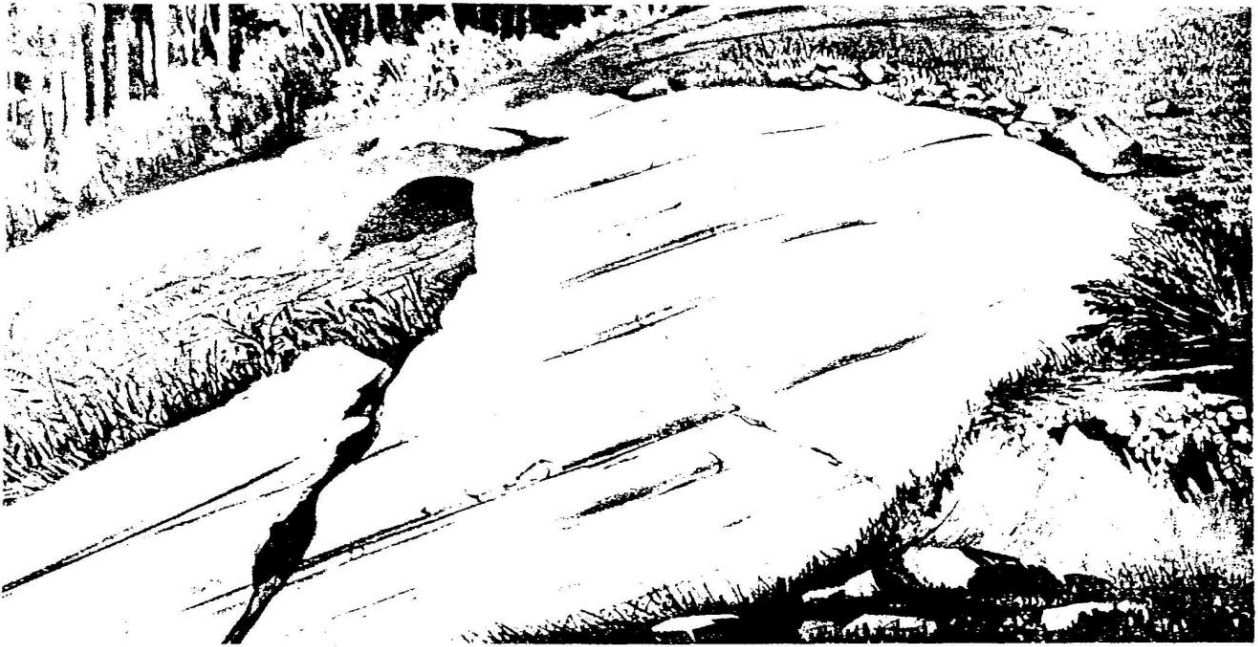

**Figure 2. Ice-polished and striated outcrop of limestone at Le Landeron, Neuchâtel, Jura Mountains (reproduction of Plate 17 in Agassiz's *Etudes sur Glaciers*, 1840).**

**Bending Ridges**

A geological map suggests that steep descents do influence swarm direction, for they show eskers and drumlins bending to the downstream direction of large river valleys. These swarm elements, however are situated not within these existing and dormant glacial conduits, but bent along side them. It appears that swarms, when still atop continental glaciers, and nearby the larger natural ice conduits, were bent by flowing ice, moving into the main downslope channel.

Good example of this is the swarms which routinely appear to have been moving in the St. Lawrence River's direction. It appears as well that this great river channel once served as a conduit for the ice sheet that covered this part of the continent. This channel would necessarily have swept the section of the swarm sitting on the ice, downstream leaving the broken ends bending towards the present river flow (See the glacial map of Canada: the St. Lawrence River). (37)

We may infer then, that valleys deep enough to carry major rivers, were, in glacial times, capable of affecting the position of drumlins and eskers that originally lay across their path. In most other cases however, where river valleys or ascents and descents of the land existed, eskers and drumlins were capable of ignoring the terrain and draped themselves across the land.

Though Grinders and Streamers alike differ in the precise mechanisms involved in sourcing and shaping an esker/drumlin swarm, present geological thought holds that the recent retreating and advancing ice sheets generated them.

In the previous pages we have documented some of the attempts geologists have made to join esker/drumlin swarms to conventional earth bound processes. Their failure is due to four critical anomalies:

  • First is the enigma of the ridge or inner core, which bears no resemblance to the product of any river or glacial valley.

  • Secondly, the roughly parallel stream beds that presumably carried these swarms cannot be found anywhere on this or any other planet.

• Thirdly, the primary core or ridge portion of a swarm is devoid of fossils, whereas the secondary mantling debris covering and filling the ridge gaps is not.

• Fourth, all ridge types have a tendency to turn to rock, some partially, some completely, a condition which cannot occur without the availability of vast quantities of cements, the earth-bound source of which is unknown.

Given the enormous intellectual capital tied up in the idea that ancient ice sheets created, moved and formed the enormous amounts of till covering the globe, a short commentary on this hypothesis is useful.

# ICE

Temperature alone is not a sufficient condition for the creation of ice sheets. An essential requirement is elevation; without it the great ice domes that presumably pushed ice sheets forward could not form. A H. Strahler and A. N. Strahler point out that Greenland, at 20° south of the North Pole but mountainous, maintains a great active ice dome while nothing but sea ice currently forms at the Pole itself. They also state that without the mountainous land of Antarctica, the South Pole would be as barren of fresh water ice domes as is the North Pole. (39)

In searching for the highlands that inaugurated and sustained recent North American ice sheets, geologists have used moraines, great arcs of till and drift, to determine the termini of our most recent ice sheets and then worked backwards to locate the elevated centers. Eskers and drumlins whose parallel ridges give the strongest sense of direction are considered the final arbiters of the origin of glacial debris. These centers, sometimes called divides, are elevated, but mountainous they are not.

**The Divides**

There appear to be three great esker/drumlin super swarms that are associated with these divides. One is in the Gulf of Bothnia in northern Europe and two are in Canada. Concerning the Canadian divides, one focus is west of Hudson Bay in the district of Keewatin —and the other is in northern (New) Quebec. These focal points are considered in traditional theory to be the location of putative ancient heights from which ice moved downslope in all directions. (Illustrations 1, 2 above and 15, 16 below) See also the glacial map of Canada. From the New Quebec divide, some diverging arms of radiating swarms of eskers and drumlins extend over a thousand kilometers into southern Quebec, Ontario and beyond. Some of the Keewatin arms at certain points appear to abut New Quebec swarms in Ontario.

**The New Quebec Super Swarm**

For purposes of beginning a in depth study of super swarms, we will focus on a significant portion of one of the two major Canadian complexes —the New Quebec swarm. Eskers/drumlins and their related forms radiate outwards from a central divide in this super swarm near the south west border of Labrador. Within this super swarm, many hundreds of kilometers from the divide in southern Ontario, is a swarm of eskers and drumlins — at least one thousand drumlins and twenty eskers — the rough centre in which lies the city of Peterborough. These eskers and drumlins trend from north-east to south-west in a band that extends from Lake Simcoe on the west, to Kingston on the east, a distance of more than two hundred kilometers. (Illustration 3, above)

**Illustration 15**

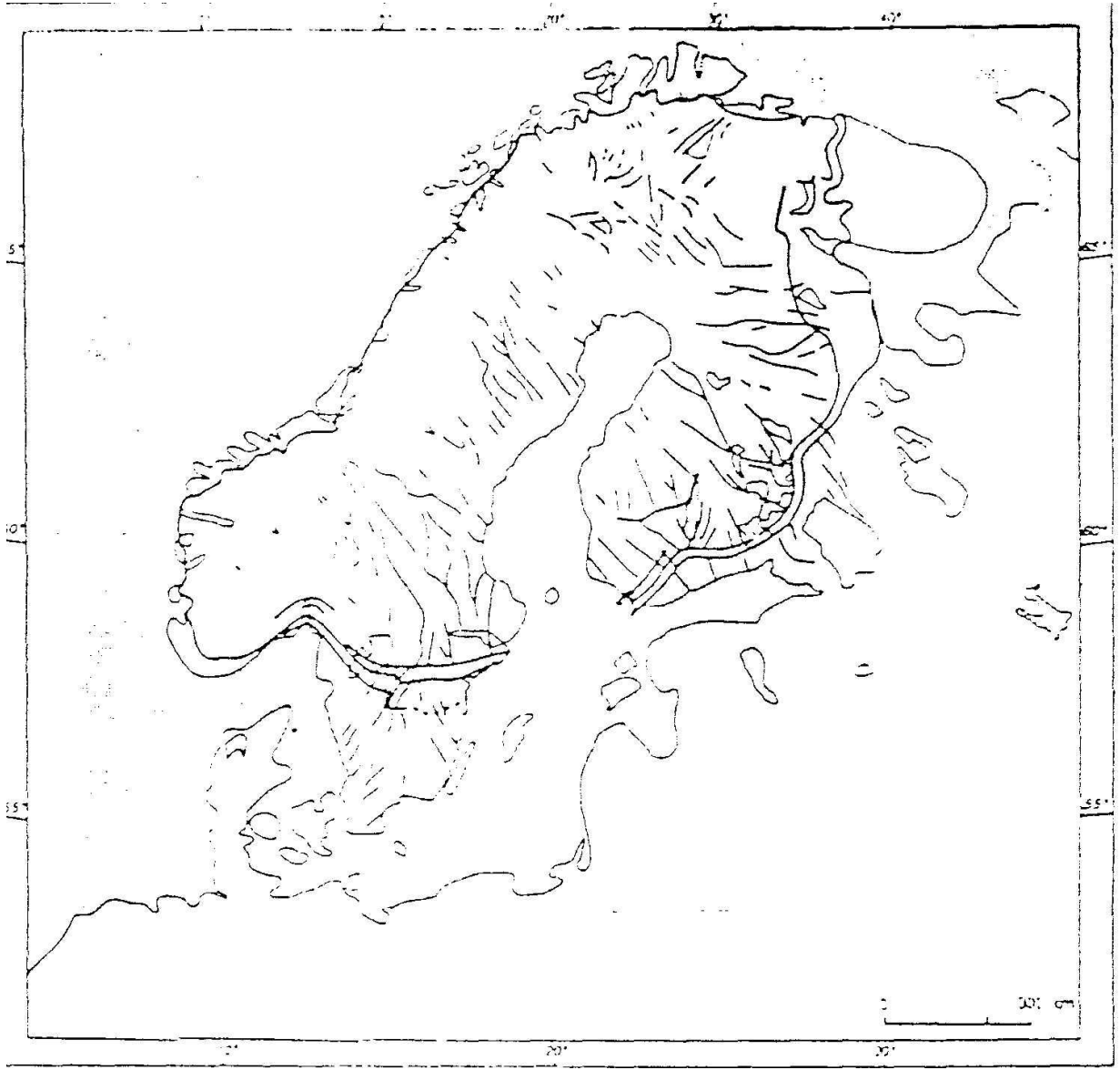

Fennoscandian moraines and main eskers of Scandinavia (compiled from general Quaternary maps of Norway, Sweden and Finland).

**Illustration 16**

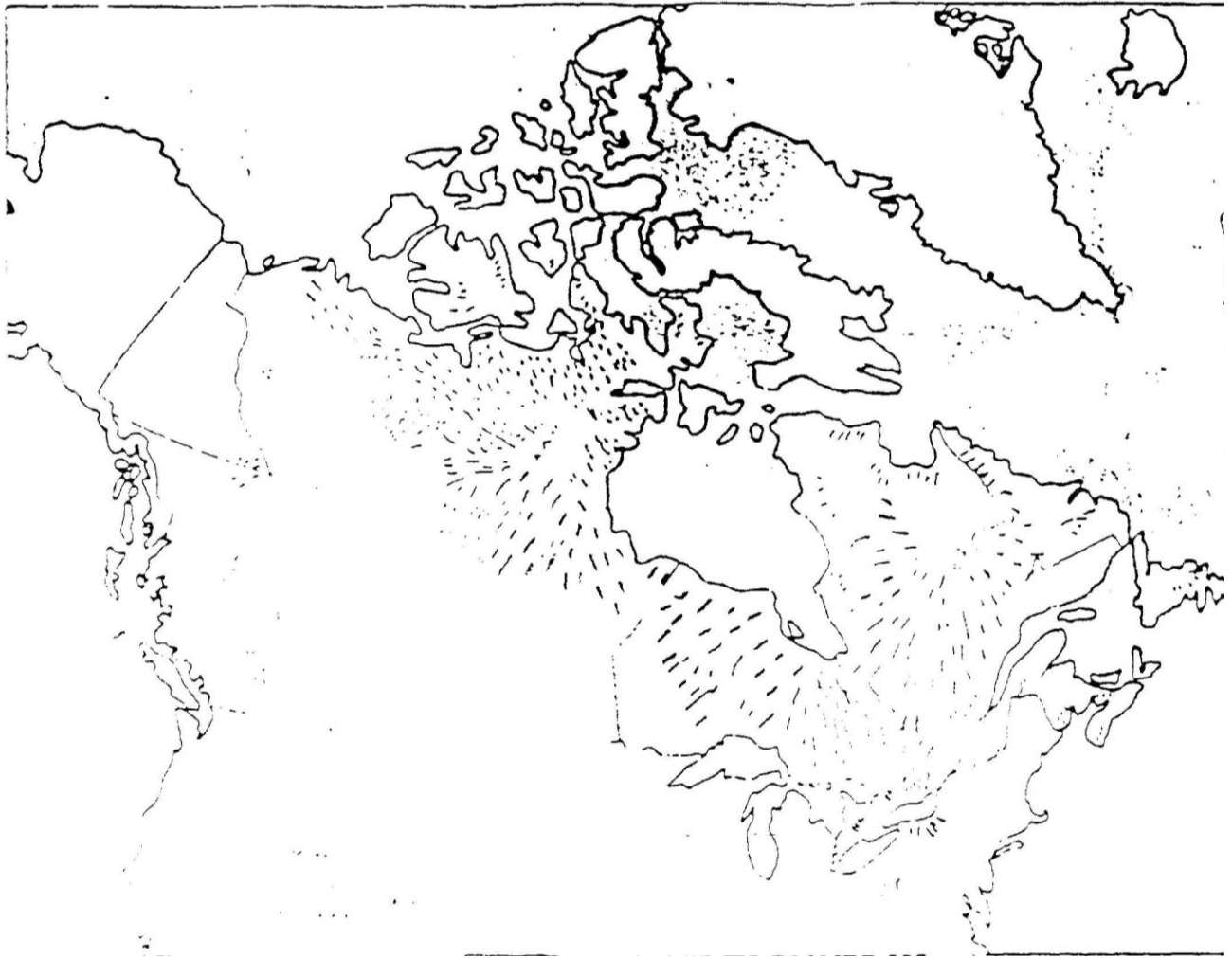

Radial pattern of eskers.

## The Peterborough Complex

A thorough examination of eskers was undertaken by McDonald and Banerjee in 1975. On behalf of the Geological Survey of Canada, they made extensive measurements of an 'esker' just north of the city of Peterborough. Located in southern Ontario, close to Canada's industrial and commercial heartland, many of the segments of the esker have been partially devoured by the demands of urban building activities. The remains —part a much larger complex, the South Central Ontario swarm – are sufficient to demonstrate the features common to an esker swarm.

## Details at Peterborough

The so-called Peterborough esker is in fact a 25 kilometer long band of narrowly separated esker ridges, trekking alongside a swarm of drumlins within the New Quebec super swarm. Many eskers in—the province of Ontario are within sight of one another, some less than 50 meters apart; the eventual merging of multiple esker strands should therefore not exceed controversy. In the Peterborough esker, this melding effect is further supported: occasional gaps and a section where two eskers travel side by side. (Illustration 3, above) Merging eskers and drumlins produce a vexing problem for both Grinders and Streamers alike, for It is nearly impossible to explain their proximity and, in some cases, their apparent fusion. In some places the water borne sediments totally mantle the closely packed ridges, forming a kilometer-wide plateau. McDonald and Banerjee note that sediments appear to come from a direction opposite to the alleged main channel. How these sediment-bearing streams could have crossed over or through one another is not explained. (39) By far the most interesting anomaly found relative to the south central Ontario band of esker-drumlin swarms is two drumlin swarms that seem to abut each other from two different directions — one representing the New Quebec Super Swarm and one representing the even more distant Keewatin Super Swarm.

Whether the Ice Rivers of the Pleistocene glaciations flowed directly south into Lake Ontario, the Trent River that flows past Peterborough, or south east through the Finger

Lakes of New York State, the direction is not that of the esker/drumlin complex, which is 45 degrees west of south. If the drumlins and eskers found in the Peterborough region were produced influenced by a traditional glacial valley, it is not apparent. The Trent River, for instance had to find its way around the esker/drumlin complex, winding its way through drumlin and esker ridges on its way to Lake Ontario.

If we used the standard method of orientation of eskers and drumlins to find the drifts origin, this path points in a north-easterly/south-westerly direction from the centre of the New Quebec swarm through Peterborough and on to Ohio. To imagine hypothetical channels and glacial rivers required to travel over the underlying rivers and mountains in order to create the Peterborough complex seems a solution born of desperation. McDonald and Banerjee may have sensed this, for they at no point in their extensive paper mention a source for the tills of the Peterborough esker. Nor is there any mention of the extent and topography of the New Quebec and Keewatin complexes. (40)

**The Swarm Enigma: A Summary**

Before turning to the cometary hypothesis, the case against an earth-based solution for esker/drumlin swarms should be summarized. Perhaps the greatest flaw in present research relates to methodology; the tendency to deal only with an esker/drumlin swarms separate elements rather than its entirety.

**Overall Swarm Structure**

To begin with, a hypothetical ground-based arterial system capable of producing an esker/drumlin swarm violates simple physical laws. For instance, neither ice sheets nor watersheds are capable of moving upslope. If ice cannot defy gravity it is obvious that it also cannot grind out the tills associated with ridges. Therefore, the eskers and drumlins found draped over hills and damming watersheds, could not have been created and transported by channels conforming to this kind of topography. This limitation confines the movement of esker and drumlin debris to hypothetical channels and tunnels high enough to move over local obstacles.

The source of these lengthy extinct arteries are the divides – centers of radiating esker/drumlin super-swarms. The great ice domes necessary for sourcing these unique arteries, however, do not currently exist. Moreover the elevated terrains necessary for their growth in a prior ice-age are also not evident.

Conceiving of an ice artery capable of laying down a single discontinuous strand in a multi- stranded swarm requires great imagination, in considering only one aspect of its function, that of transportation, the artery must be capable of trending hundreds of kilometers. Tracking roughly parallel to its swarm companions, over intervening hills and valleys , and then, from time to time, apparently sealing itself, interrupting the deposition of the characteristic ridge complex, and continuing the new dual-regime segments further along the swarm's path. In addition to this sophisticated feat of transportation and laydown, this ice tunnel or super-glacial channel must perform other necessary tasks such as grinding gathering, lifting and sorting. It is therefore understandable that no attempt has so far been made to explain the esker/drumlin complex in its entirety.

Regarding the specific claims geologists have made for valley glaciers as generators of what they refer to as eskers, our review of the evidence found this approach unpersuasive. The debris gathered and deposited by valley glaciers bears little resemblance to the drift from which esker/drumlin ridges are constructed.

Moreover, no valley glaciers, past or present have produced eskers and drumlins having:

      a) Classical dual regime features radiating from a divide
      b) Parallel ridges
      c) Multiple ridges, which converge, climb and cross over one another
      d) Ridge lithification
      e) Unsorted fossil-free debris

No earth-based solutions appear to have been advanced by geology that provides a satisfactory explanation of the origin of esker/drumlin ridges.

# THE COMETARY HYPOTHESIS

Much of the argument thus far has been indirect;  the surficial deposition of ridge systems lying upon, but not disturbing underlying strata; their tendency to create their own unique, dominating architecture independent of controlling topography such as watersheds and hills; the sterile nature of ridge interiors, unsorted and unstratified by running water, yet mysteriously concretized in part or full. None of these characteristics of ridge systems is explained by prevailing glacial theory.

## The First Cometary Connection

The first modern investigator to notice the anomalies of the drift and to connect them to comets was a remarkable American by the name of Ignatius Donnelly. Though he never mentioned eskers or drumlins, he used the drift as the principal evidentiary tool in linking the tail of a comet to vast volumes of rock debris.

In 1882, Donnelly wrote one of the best-sellers in the United States: <u>Atlantis, The Antediluvian World</u>, which presented a new approach to the legendary lost continent of Atlantis.

## Ragnarok

Encouraged by the critical and financial success of his first literary effort, Donnelly wrote a sequel examining the cause of the island's destruction and proclaiming its present location. <u>The Destruction of Atlantis, Ragnarok: The age of Fire and Gravel</u> (41) claimed that the instrument of catastrophe was no mundane phenomenon, such as an earthquake or invasion, but the earth's encounter with the tail of a great comet, which, in the course of destroying Atlantis, covered most of the globe with its debris. Donnelly traced the comets fallout directly to the drift.

He described this material in his effective folksy prose: "Upon the top of the last of this series of stratified rocks we find the drift. What is it? Go out with me where yonder men are digging a well. Let us observe the material they are casting out. First, they penetrate through a few inches or a foot or two of soil. Then they enter a vast deposit of sand, gravel and clay...

… It may be 50, 100, 500, 800 feet before they reach the stratified rocks on which this drift rests. It covers entire continents. It is our earth. It makes the basis of our soils. The railroads cut their way through it, our carriages drive over it. Our cities are built upon it, our crops are derived from it. The water we drink percolates through it. On it we live, love, marry, raise children, think, dream and die, and in the bosom of it we will be buried."

Donnelly carefully analyzed both approaches to the problem of drift development and rejected them as incomplete. Neither flood nor ice could lift the till - the peculiar sorted rock debris found in previously glaciated terrain - to the tops of the highest mountains. Neither could these agents account for their production, order and distribution.

**The Cosmic Grinder**

Donnelly's principal insight was that drift, the world's extensive deposits of rocky debris, sand and clay, was the product of a unique grinding process found in the tails of large comets. Most drift, particularly the rocky till strewn around the landscapes of many countries was in Donnelly's day considered to be principally the product of advancing and retreating ice sheets. Contemporary geologists, however, had already noted some inconsistencies relating to the theory. "There is something very peculiar about the shape of the stones," said the highly respected James Geikie. "They are neither round or oval like the pebbles in river gravels or the shingle of the seashore, nor are they sharply angular like newly fallen debris at the base of a cliff, although they more closely resemble the latter than the former. They are indeed angular in shape but the sharp corners and edges have invariably been smoothed away. …Their shape, as will be seen, is by no means their most striking peculiarity. Each is smoothed, polished and covered with stria or scratches, some of which are delicate as the lines traced by an etching needle, others deep and harsh as the scores made by the plough upon a rock. And what is worthy of note, most of the scratches, coarse and fine together, seem to run parallel to the longer diameter of the stones which, however, are scratched in many other directions as well." (42)

Geikie also noted the difficulty of tracing the till to existing glacial valleys: "...No till or drift is now being formed by, or under, the glaciers of Switzerland. Nevertheless, till is found in that country disassociated from the mountains…" (43)

**Comet Debris Credible**

Donnelly not only rejected ice as the creator of till, he attacked the essential mechanism of erosion as the producer of sediments. He pointed out that the tail of a large comet was not only suited for making the larger fragments of till but also for segregating particles into the anomalously segregated clay deposits.

In the authors words, comet "debris arranges itself in a regular order, the largest fragments are nearest the head, the smaller are farther away, diminishing in regular gradation until the farthest extremity. The tail consists of sand, dust and gasses. Through the continual movement of the particles of the tail, operated on by the attraction and repulsion of the sun, the fragments collide and crash against each other. By natural law each stone places itself so that its longest diameter coincides with the motion of the comet. Hence, as they scrape together they mark each other with lines or stria lengthwise of their longest diameter. The fine dust ground out by these perpetual collisions does not go off into space but pack around the stones. But still governed by the attraction of the head, it falls to the rear and takes its place like the small men of a regiment in the farther part of the tail." (44)

Donnelly's mechanical/gravitational hypothesis for explaining till striations and particle separation of sediments is open to criticism. The longitudinal sizing of tail dynamics has yet to be observed in comet tails and the larger to smaller sizing not reflected along the axis of esker/drumlin swarms nor in any other drift environment. His observations concerning the magnetic properties of clays, however, are another matter.

## Magnetic Properties

Clay, which is a wide classification, covering all heterogeneous finely ground up rock, is inexplicably organized into iron rich and iron poor beds. It was already known in the author's day that the color of clays was due to this peculiarity. Mica and hornblende contains considerable iron oxide, whereas feldspar yields only a trace, or none; therefore, clays that derived from feldspar are light colored or white, while those partially made up of decomposed mica or hornblende are dark, either bluish or red.

Donnelly observed: "The particles ground out of feldspar are finer than those derived from mica and hornblende, and we can readily understand how the great forces of gravity might separate one from the other, or how magnetic waves passing through the comet might arrange all the particles containing iron by themselves and thus produce their marvelous separation of the constituents of the granites which we have found to exist in the drift clays." (45)

## Donnelley's Contributions

Donnelly did not have access to spacecraft observation of Halley and other comet nuclei. He did, however, present the first modern argument for the cometary origin of drift and plausible extra-terrestrial explanations for till-related enigmas.

He noted the unique nature of the drift and its propensity to develop into unusual concretions such as the hardpan and roche moutonnee. (46) He also pointed out the unfossilized condition of drift as well as the occurrence of clay beds organized according to iron content. (47)

## Donnelley's Comet

Ragnarok's comet has been contemplated more by catastrophist literature and cinema than by modern astronomy, and most astronomers would argue that we have no evidence that comets grow to planetary size. Nonetheless, comets, as cosmic phenomena, have always been in a class of their own, their appearance prior to the age of Newton, a bad omen. Donnelly's contention that a single comet could cover the entire earth with its debris was, in any era, an unsettling idea.

## Convulsions and Catastrophes

Donnelly painted a bleak picture of a comet's destructive power, which ended the "Golden Age" at some indeterminate time. (48) The drift, he wrote, "marks probably the most awful convulsion and catastrophe that have ever fallen upon the globe". (49) Accompanying these deposits of "continental masses of clay sand and gravel" were convulsions which produced "cracks and fissures which reached down through many miles of the earth's crust to the central fires and released the boiling rocks imprisoned in its bosom". (50) He cited the Grand Canyon and Scandinavian fjords as examples of formations wedged apart by the action of cooling magma in gaping tears in the earth's crust.

This graphic image unwittingly describes the shearing effect the laydown of great mountain ridges would have in extruding igneous rock from beneath thinner sections of the earth's crust. The Atlantis disaster, according to Donnelly, was not the only encounter the earth had with comets. "We find that all through the rocky record of our globe, the same phenomena which we have learned to recognize as being peculiar to the Drift Age are, at distant intervals, repeated. The long ages of the Paleozoic time passed with few or no disturbances, the movements of the earths crust oscillated at a rate not to exceed one foot in a century. It was an age of great peace. Then came a tremendous convulsion."

## Monstrous Luminaries

Drawing on the mythological and physical record, Donnelly posed a question only now being asked by astronomers. "Can it be that there wander through immeasurable space upon an orbit of such (great length) that millions of years are required to complete it, some monstrous luminary so vast that (its tail) showers down upon us deluges of debris while it fills the world with flame?"

Donnelly was confident that analysis of comet debris would be a key factor in indentifying the errant giant comets of the past. He was mistaken in gauging the interest of future scientists; contemporary professionals merely ignored Donnelly but were brutal to his successor. Immanuel Velikovsky (1895-1979), attempted to bring greater precision to the theory and chronology of catastrophism. His best selling Worlds in Collision (1950) brought Donnelly's generalization into a much more detailed cosmic scenario. Donnelly's great comet became, through Velikovsky's intensely researched work, the 'protoplanet' Venus. This nascent planet, according to the writer, came perilously close to the earth in the middle of the second millennium BCE causing global upheaval and dislodging the orbits of Mars and the Moon before settling into its present peaceful orbit.

Although Velikovsky believed the encounter with Venus involved the earth's acquisition of organic and inorganic debris, he gave Donnelly's insights on the drift an insufficient and undeserved dismissal. (51)

Velikovsky confined his speculations on recent mountain building to crustal deformation. These upheavals of the earth's surface were triggered by the close approach of Venus and Mars, which dislocated the earth's terrestrial axis. Velikovsky made no special mention of the drift.

**Asteroids Embraced**

The attitudes of mainstream science to extraterrestrial-caused global upheavals have changed markedly from the days of Donnelly and Velikovsky. Neo-catastrophists have now embraced asteroids and comets as the agents of explosive geological and biological change. The U.S. congress has gone as far as appropriating funds for an asteroid space watch. Yet direct evidence of the impact of comets and asteroids has proved elusive. So-called impact craters, for instance, have yet to reveal, within their capacious raised rims, significant traces of meteoric debris. Mathematical theory has been employed to explain the absence of any vestige of the asteroid due to its vaporization. (52)

**Ideal Solution**

An ideal extraterrestrial solution to the enigmatic structures connected to the drift would account for both overall structure and internal make-up.

In the following postulation of a cometary encounter, no precise quantitative data such as dust tail payload or number of laydowns will be provided.

………………What If?

In 1995, a fragmented comet – 14 fragments – each with its own tail - impacted and disappeared beneath the thick cloud cover of Jupiter. Ground and satellite-based observations measured significant electro-magnetic activity prior to and incident to contact. (53) The angle of descent was approximately forty-five degrees.

What if a somewhat similar encounter occurred with our planet with the following qualifications:

1. The earth, instead of capturing a comet's nucleus, would only capture a segment of its tail, or in the case of multiple comets, segments of their overlapping tails.

2. The event was divided, somewhat arbitrarily, into a mountain-building and drift (esker/drumlin)-building phase.

3. The orbit of both phases generated a gentler landing angle - less than twenty-five degrees.

Whether the great comet, conjectured as the mountain- builder, created the Cordillera, the mid-oceanic ridge system, the Appalachians or all of them, is not known. Conceivably, all major mountain chains could be the result of separate encounters or multiple returns of a single giant. This issue will be discussed in a later paper.

## Common Characteristics

Whether undersea or continental, mountain ranges possess the same general characteristics. Seen from a great enough perspective, they model, in much larger scale, the esker/drumlin swarms we have examined in this paper. Their great mass however, has removed them as ice artifacts. Neither the earth's atmosphere nor the landing cushion of oceans and snows would have had any great effect on their final laydown pattern. On the contrary it would be these gigantic ridges that would imprint and impose their laydown pattern and chemical makeup on the topography, climate and finally, the biosphere of this planet.

## Pattern of Orbital Debris

The rough parallelism and discontinuities found in both mountains and esker/drumlin systems is often the point of departure in attempting to explain their genesis. Aside from their massiveness, mountain ranges tend to have one main axis, whereas drumlin/esker swarms may radiate a number of axes from a common centre. (Illustration 1, 8 above and 17, 18, 19, 20 below) In most other respects, they are microscopic replicas.

**Illustration 17**

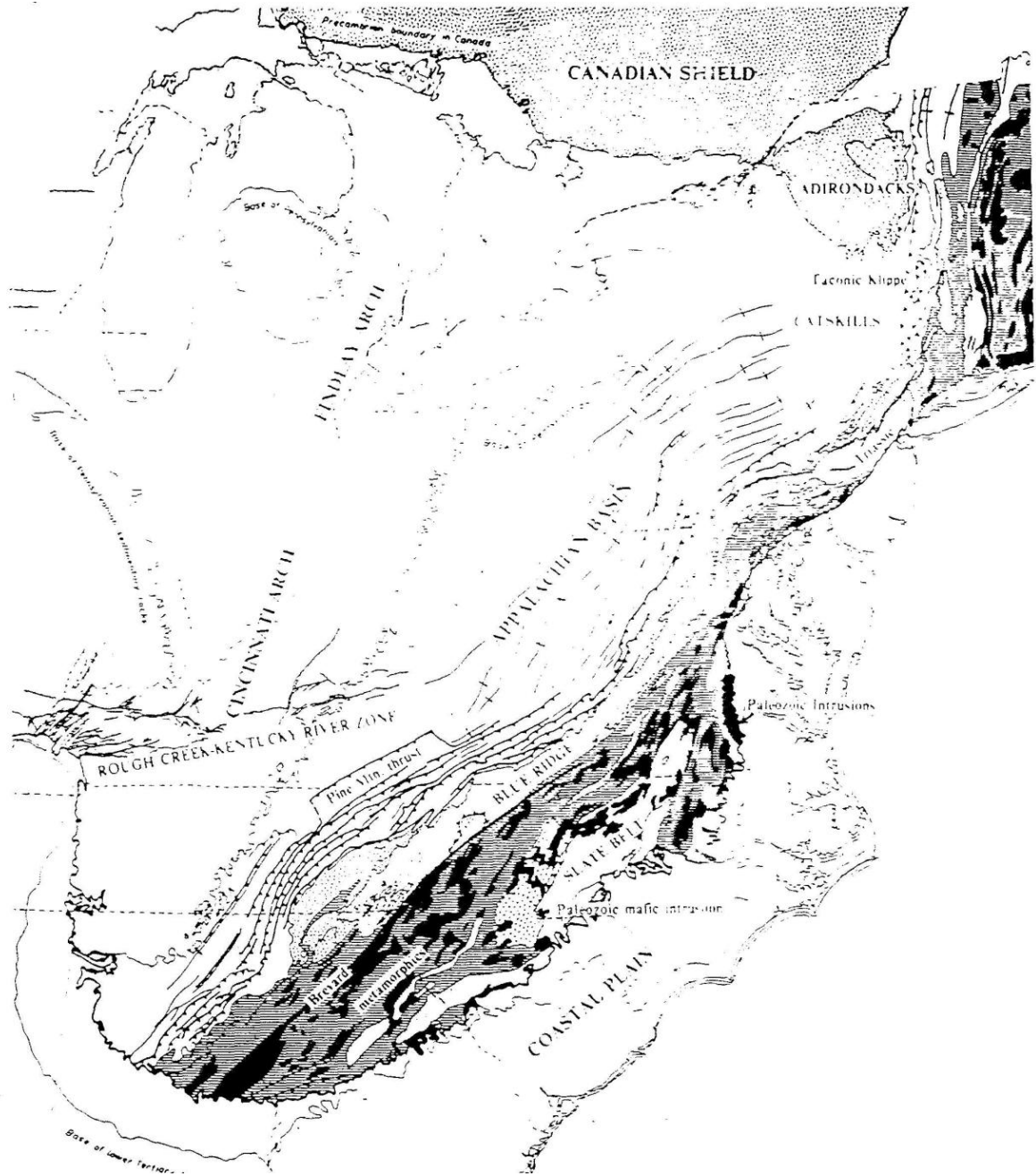

**Tectonic sketch map of the southern and central portion of the Appalachian Orogen. (After the American Assosiation of Petroleum Geologist Tectonic Map of the United States,1969)**

**Illustration 18**

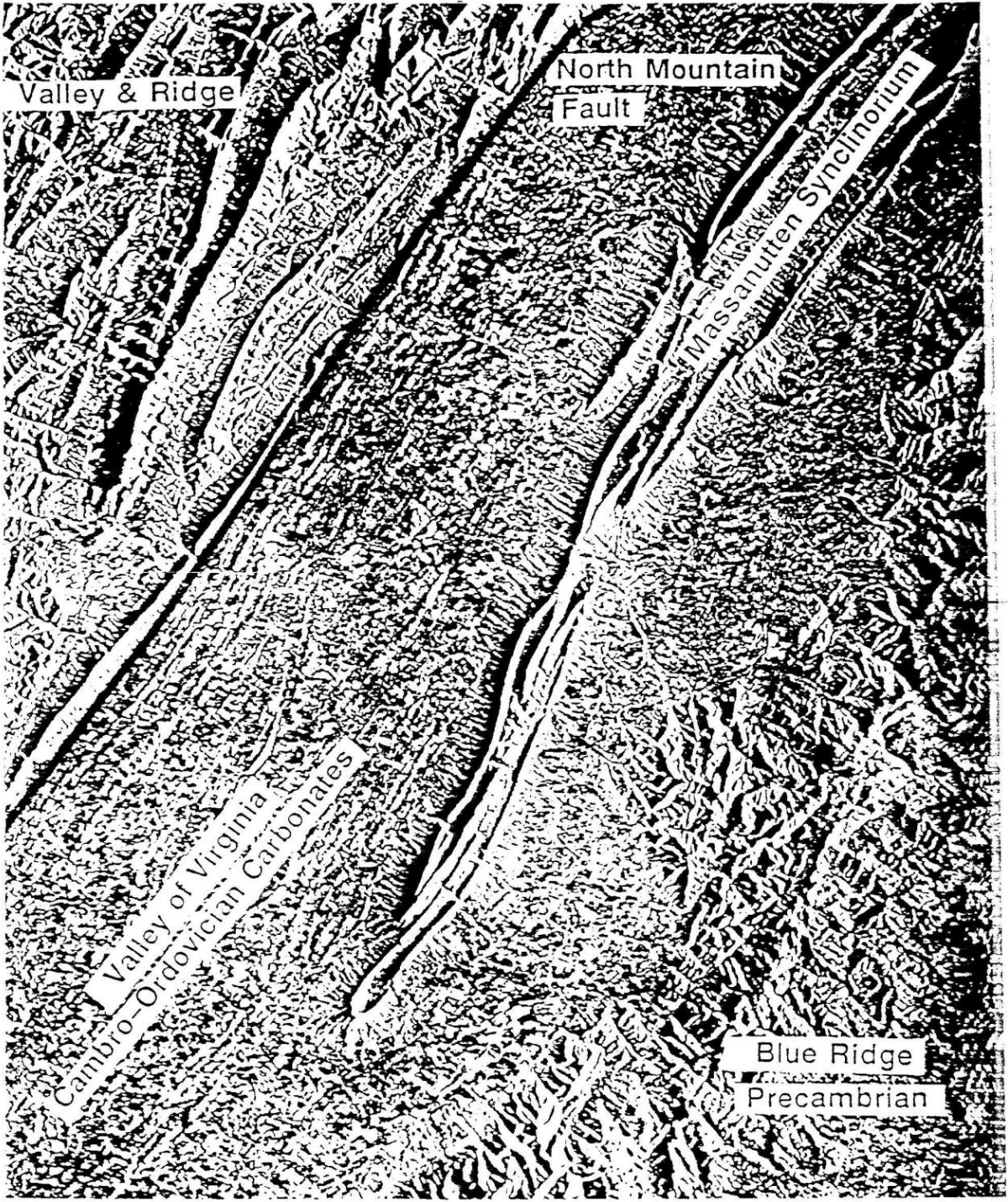

Side looking radar image of a portion of the Blue Ridge, Great Valley, Massanutten synclinorium, and the eastern edge of the Valley and Ridge.

## Illustration 19

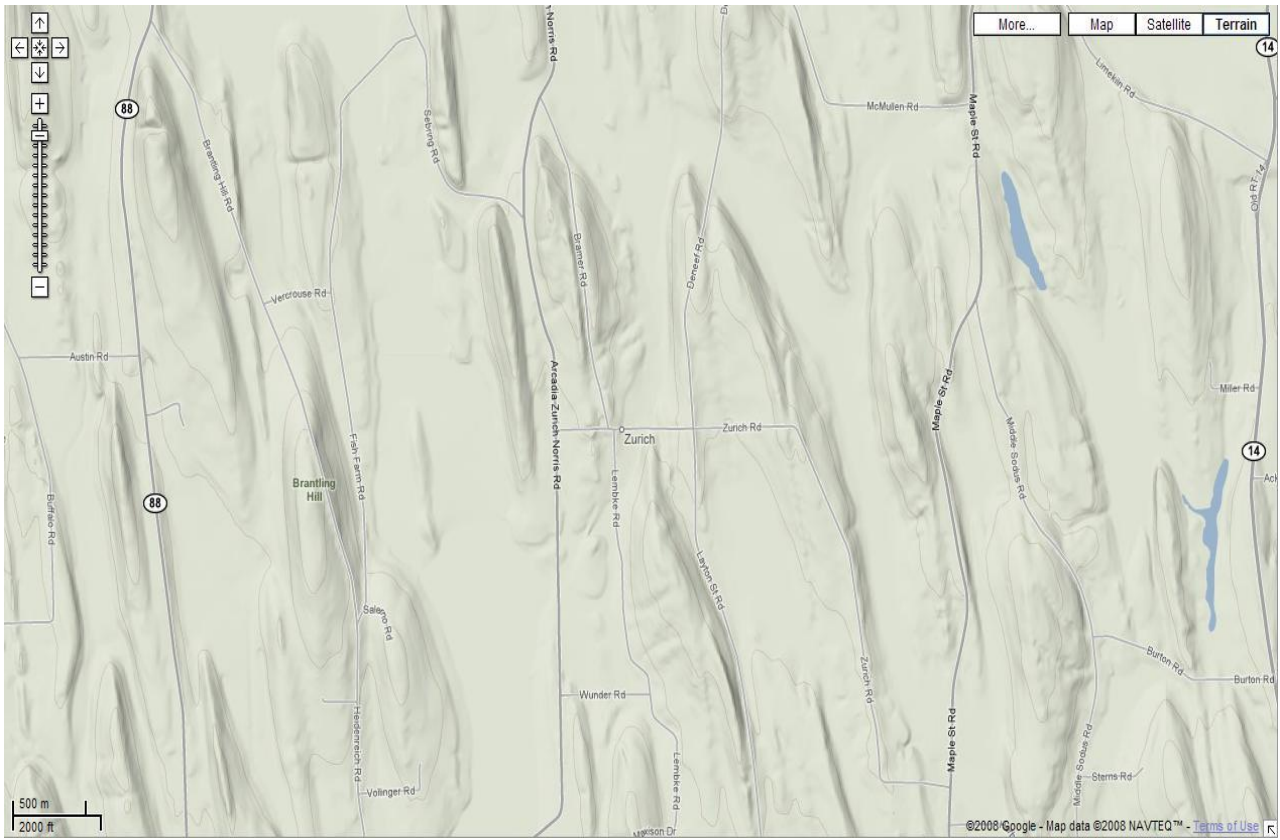

A swarm of drumlins in Ontario, Canada, formed by ice moving from upper left to lower right. As drumlins go, many of these are unusually long and narrow, but the classical blunt upseam nose and tapering narrow tail are apparent on most. Photo taken using Google Maps terrain view.

**Illustration 20**

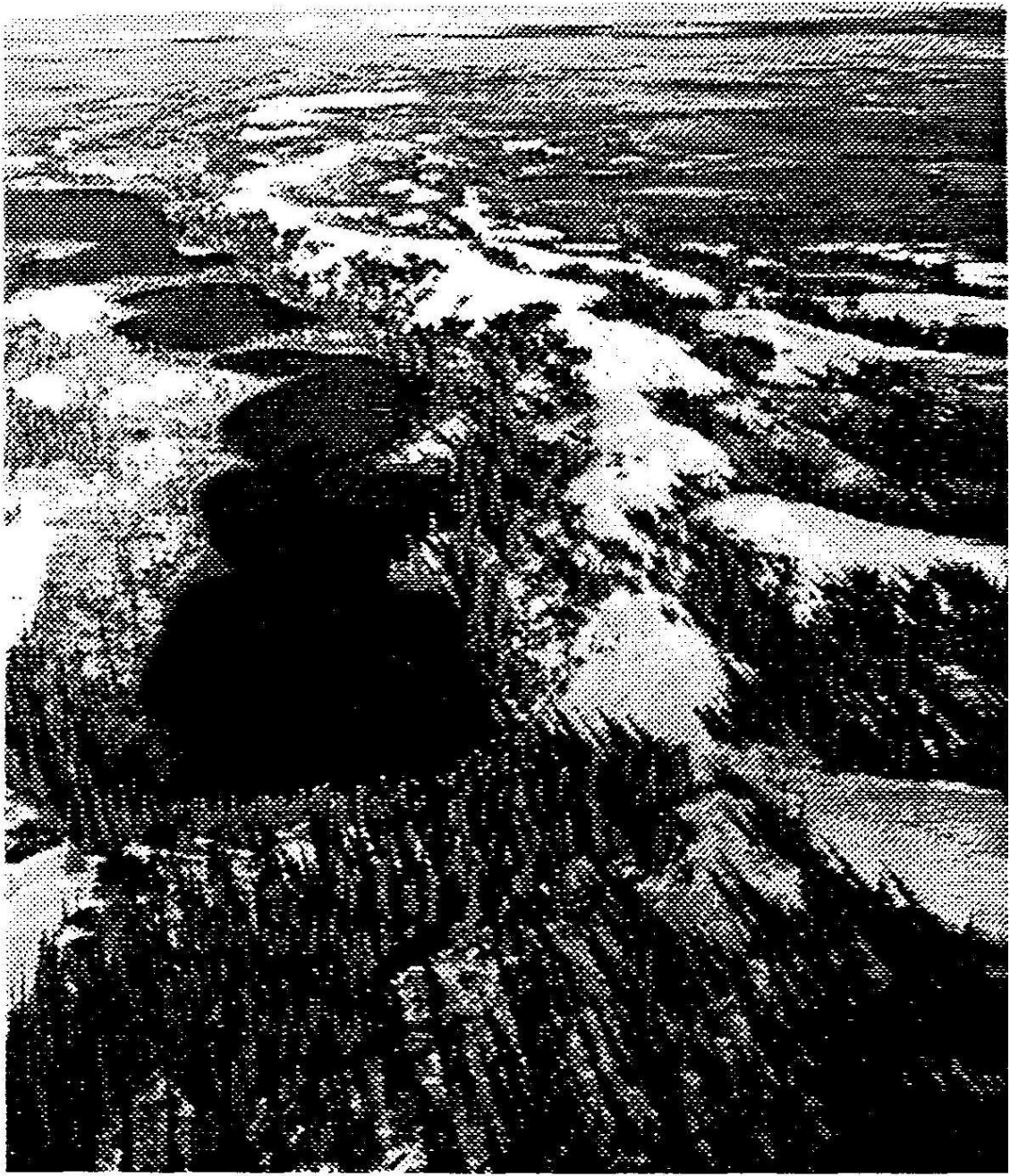

On the southern edge of the Barren Lands, where trees give way to open tundra, an esker, revealed by its sandy surface, stretches towards the horizon. Often the only dry lands in warm weather, eskers are travelled by man and beast.

**Original Hypothesis**

These similarities compel the writer to hypothesize an original mountain-building event followed later by a return of the reduced and fragmented giant, which created the drift. Before embarking on this cosmic excursion, critical aspects of how cometary debris is distributed to its dust tail should be discussed. (Illustration 21 below) The best evidence connecting comets to ridge systems is disclosed by the manner in which cometary debris is dislodged from and distributed to its debris tail. Despite the close observation of Halley and other comets by satellite and ground-based telescopes, the process is little understood and sketchily described.

**Debris Distribution**

Each ridge may be traced to a single stream of cometary debris called a jet, discharging intermittently from a comet's nucleus. A comet's debris tail is made up of a series of jets spreading along the comet's orbital plane in a band of ribbons. It is this unusual attribute of comets that accounts for the unique patterns of ridge systems on earth.

The separated and intermittent quality of jet emission would account for both the individuality of ridges and their discontinuities. But what of the ridge profile itself? Jets, while concentrating the ejecta from the nucleus, appear to be hundreds of meters across even on a nucleus as small as that of Halley's Comet. (54)  How does this translate into a crested ridge that is sometimes only a few meters wide? More troubling is the relatively tight packing found in the swarms themselves. Some eskers and drumlins are so close together they ride, up the sides of their companions. The same arrangement is apparent in the Appalachians where the ridge structure is so closely defined as to appear groomed by some gigantic rake. (Illustration 17, 18 above)

**The Jet Anomaly**

After ejection from the nucleus, the development of the shape and orientation of jets represent the most important of the anomalies associated with its debris tail. It is a little known fact that a comet discharges its debris along the plane of its orbit. William Corliss, in his catalogue of anomalies quotes McCrosky in the <u>McGraw-Hill Encyclopedia of Science and Technology</u>: (55) "In all cases the tail material is closely confined to the plane of the orbit. In comets that show more than one type of tail which show type 1 streamers, the various structures disappear when Earth passes through the plane of the comet's orbit. It is as though one were looking at a sheet of paper edgewise."

This ejection pattern effectively explains why comet tails often resemble fans. (Illistration 8, above) It also gives a satisfactory explanation to other peculiarities of the lesser ridge systems.

**Illustration 21**

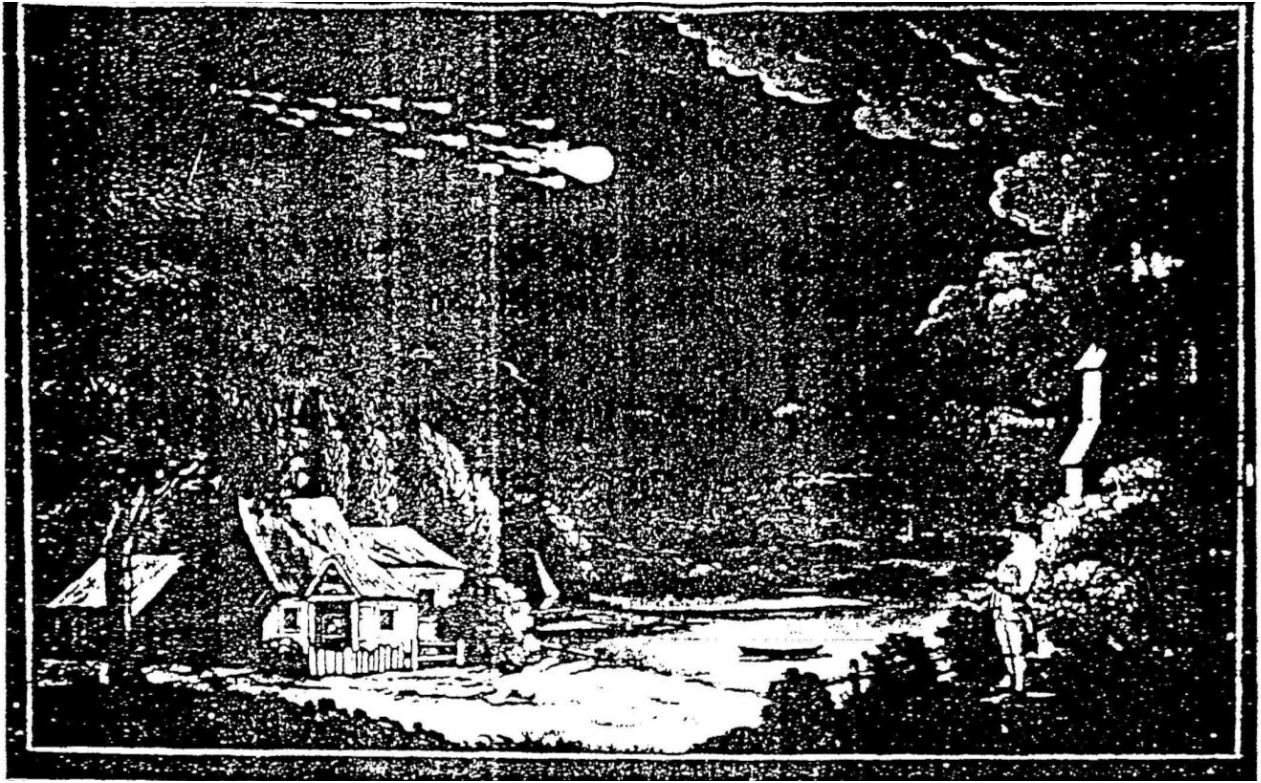

The handwritten legend beneath this picture states that it is an accurate representation by Henry Robinson, Schoolmaster, of the meteor of August 18, 1783, as it passed over Winthrop, near Newark upon Trent. At first it appeased as one ball of fire, but in a few seconds it broke into smaller ones. This meteor, said Robinson, is of the species with the great Physiologist Dr. Woodward and others call *Draco Volans* or *Flying Dragon*. The fireball was so spectacular that several observers published descriptions of it in the next few months, but no meteorites were collected in connection with the event. (Courtesy of the British Museum of Natural History).

**Ribbon Transfer**

We have already looked at the fact that a comet's dust tail is made up of a series of jets spreading along a comet's orbital plane in a band of ribbons. The following hypothesis assumes that a segment of a multi-ribboned tail made a minimum series of three separate touch-downs, and that these landings account for the radiating patterns of drift found after the retreat of the latest continental ice sheets.

A series of flattened ribbons following the orbit of a comet would, in transferring part of its tail to the earth, continue to orient its new orbit around the earth in the same aspect – that is to say, with the tail ribbons maintaining their orbit around the earth in the same aspect that rings orbit around Saturn.

The debris contained by each ribbon would therefore land perpendicular to the earth's surface like a long blade making a shallow cut into a frosted cake. Each vertical ribbon could land, depending on a number of factors, on top of another, which would account for the tight packing of ridge systems.

**Ridge Mantle Division**

The division into ridge and mantle debris would be due to grading found in each ribbon. The lighter particles, more easily ejected from the comet nucleus, would reach the furthest position (on the ribbon) from the centre of the nucleus. This gradation, though gradual, would eventually divide on contact with the earth's atmosphere into two regimes determined by particle size and mass. The outer regime would contain the clays, silts, water crystals and organic hydrocarbon gases. The inner regime would contain the boulders, cobbles, gravel and sand.

**A Mega-Model**

Before examining esker/drumlin swarms in detail, the ancestral comet that created the earth's distinctive basin and range topography should be briefly alluded to.

Though not the focus of this paper, these great ridge systems, with a few important exceptions serve as an enlarged model of the lesser ridge swarms examined in this paper.

**The First Encounter**

The first encounter would involve the capture by the earth of a substantial portion of a planet-sized comet. Assuming the number of jets to be independent of the size of a comets nucleus, and an active nucleus to contain 17 - the number observed on Haley by satellite (56), the pattern of the Appalachians would not be disproportional for a mega-comet's payload. Seventeen ribbons of debris landing at an average of ten kilometers per ridge unit would create a chain one hundred and eighty kilometers wide. This width is not inconsistent with Appalachian ridge structure. A thirty kilometer square section provides an idea of the proportion of a well-defined mountain ridge system. (Illustrations 8, 17, 18 above)

Orogeny or mountain building is not the direct subject of this inquiry, but the cometary hypothesis could shed light on the enigma geologist's face when looking for an endogenous explanation for the thick strata of salt, phosphate, metal precipitates and coal found adjacent to the great ridges. A cometary hypothesis would explain them.

**Filling the Great Basins**

In the first great encounter, debris-laden floods, mixed with earthly and cosmic waters, surged through the valleys and pool in the plains, their course determined by freshly-laid down, giant, roughly parallel ridges. The intervening valleys and plains became capacious basins—oceans, great fresh-water lakes, navigable rivers. These bodies of water, as well as sedimentary deposits and glaciers, were defined and contained by these great mountain ridges.

Enormous quantities of inorganic and organic suspensions and solutions settled and precipitated into the declivities. Distinct layers of silt, coal, limestones, shales became stacked on one another in strata thousands of meters in thickness and continental in breadth. These strata, called bedding planes, sometimes rose to the top of their basins and became great plateaus.

**Tearing the Crust**

The most important difference between the ancestral comet and its fragmented descendants was the great force developed by its massive jets. The jets of the ancestral comet were so great they literally tore open the earth. Fragile crustal surfaces, sheared by the great weight of the recently-implanted ridges caused molten rock to extrude through the tears, creating great lakes of frozen igneous rock in continental and sub-oceanic basins. More localized imperfections in the crust responded to the sudden pressure by forming volcanic chains along mountain belts.

Critical features of the great encounters were the craters formed by enormous discharges between the earth and the impinging debris. Referred to in other work, these craters, now assumed to be the result of physical impacts are essentially electric discharge phenomena.

**Landing a Swarm: Part One**

Apart from the drama of earth-splitting mountains, the model of a dual regime of heavy and light debris serves both mountain and esker alike as the landing pattern of cometary tail jets.

We have no knowledge of the velocity of the particulate matter that created this planet's ridge systems. The general patterns of the drift, however, suggest that they come out of an earth-circling orbit. That ribbon-like rings may have orbited the earth is suggested by Saturn's distinctive rings. The problem of a landing speed capable of capturing and preserving ridge systems is not known. Meteorites are theoretically capable of impact at speeds greater than forty kilometers per second, (57) enough to possibly vaporize the missile. On the other hand meteorites have been known to fall out of the sky and land in such a shallow grave that they can be lifted by a single scoop of a shovel. Smaller bits of meteoric debris have been collected on the waxed wings of U2s and casually picked off the surface of Antarctic snows. The water from thousands of daily ice-comet impacts, according to a recently-verified hypothesis of Louis Frank have apparently filled the earth's basins with its entire water supply and continue to do so. (58)

In the early part of this century, a parade of fireballs, presumably accompanied by a substantial meteoric shower, passed by southern Ontario at such a leisurely rate that they were visible for three minutes before they disappeared beneath the horizon. (59) In other words, it is possible that orbiting debris, given the physical evidence available on known meteoric falls, suggests that a jet could land on a cushion of ice and snow and retain its essential character.

**The Landing, Part Two: Getting the Drift**

Like bullets shot into a box of cotton batting, the uncountable particles of each cometary ribbon burrowed into the ice. These ribbons became separated and filtered by the earth's atmosphere. The post-depositional movements of the underlying ice and its later melting altered the original laydown pattern. Nonetheless, the essential landing pattern is still clearly discernible in the present artifacts of continental drift.

In northern Canada in the district of Keewatin, in northern Quebec and in Europe, centered in the Gulf of Bothnia, are three radiations of eskers, drumlins and their related ridge structures. (Illustrations 1, 2, 15, 16 above) Their artifacts are primarily represented by esker/drumlin swarms and concentric arcs of moraines. The greatest moraines appear to gather on the furthest reach of the last great ice sheet. We are now in a position where glacial theory began in the nineteenth century i.e. the drift has to be delivered to the top of ice sheets in order to eventually account for its subsequent arrangement of moraines and ridge systems.

**Shoemaker-Levy**

Comets, as they make their journey around the sun have been observed breaking up. Shoemaker-Levy recently left 15 fragments on Jupiter (60), each with its own debris tail. There are over 100 ridges in the extended esker radiations of the New Quebec super-swarms along. As many as 50 fragments could have been involved, each with its own multi-jetted tail.

We are assuming that the tails captured into an earth orbit resembled a large segment of an elliptical banded disc. Each band would be composed of a great number of ribbons, with each ribbon originally attached to a single fragment. Each fragment, in turn, would have its own multi-jetted tail. This tail would form a flat beribboned series of bands because it would obey the same laws governing the unusual payoff of jets from a comet's nucleus. Each time a nucleus is fractured, it is assumed that the smaller fragments, like any particle of dust, would part from its parent or twin on the plane of the fragmenting comet's orbit.

**Different Chemical Locations**

Different fragments may represent different chemical locations on the nucleus and would probably fragment unevenly. We should therefore expect that each series of - ribbons issuing from a distinct fragment would display different qualities.

This hypothesis is supported by the variety of ridge patterns found in super-swarms. Some sub-systems in a super-swarm show collective features that compel geologists to consider them separate geological phenomena--esker, drumlin, rock drumlin etc. The Peterborough complex, for example, contains its compound esker within a much larger drumlin swarm. Are these distinctive units the product of different fragments? Some sub-systems do display different ridge mass, thickness of mantle and inner chemical makeup. They also appear to have greater and lesser separations. In the district of Keewatin, eskers range in parallel paths in 13 to 14-kilometre wide ranks in a swarm. (61)

**Creating a Divide**

The organization of jet ribbons on a flat plane not only explains the parallelism, discontinuities and individual nature of swarms, it also provides a solution to the question of the origin of the divides.

**Stacked Up for Radial Landing**

It is appropriate to think of the multiple tail segments orbiting the earth as a cosmic rainbow. Each ribbon of the 'rainbow' became detached from its perpendicular stacking order, landing sequentially as gravity pulled each jetted ribbon to its ice-bound landing zone. Each touchdown would be delayed by the distance between each ribbon: resulting in the distance between each ridge formation. Each second of delay between touchdowns would move the earth approximately one thousand meters per second at the latitude of Peterborough. Each jetted tail would form an additional spoke in the radiating pattern. Each successive ribboned jet would land at an angle determined by the vertical distance between the remaining stacked orbiting ribbons.

**The Landing**

Each landing would occur at a different meridian of the earth's diurnal rotation, creating the divides mentioned above. The jets would cross over one another and radiate creating the pattern of a divide.

**The Right Sort**

Esker interiors demonstrate only one sign of sorting. Till deposits within esker/drumlin edges show limited organization of the fragments, which is that large numbers of stones may lie with their long axes parallel to the flow direction of the glacier. It is apparent that the initial phase of esker development involved the passage of its rocky debris through some kind of medium. Was it our own atmosphere?

In an esker's inner core, each particle becomes smoother as it increases in size. Examples of this progressive ablation are found in the Peterborough esker.

This effect is strikingly similar to the solution found for the first astronauts. The Mercury capsule was guarded by a shield fabricated from a composite of rock dust and adhesive. As the capsule touched the atmosphere, the enormous heat of re-entry friction was dissipated by the sloughing off or ablation of the top layers of the shield. The same the same process which was at work in the penetration of each jet segment as it entered the earth's atmosphere. From larger boulder to microscopic clay particle, the larger component of this closely-packed shower protected its smaller trailing companion by deflecting the oxygen-laden atmosphere- the larger particles becoming progressively burned away by the unforgiving atmospheric friction.

## Sorting the Anomalies

Atmospheric sorting also explains other ridge anomalies. The softer rocks such as schistose should show a declining presence in the esker as they dissolved and broke up moving down hypothetical channels. But this doesn't happen. Schistose and other soft rocky debris are equally represented in all points of an esker/drumlin ridge, independent of distance from either end. (62)

## Ridge and Mantle

The passage of the contents of an esker/drumlin swarm through our atmosphere accounts in part, for the two regimes of ridge and mantle. Some eskers are almost devoid of mantling materials, indicating that much of the silt and clay particles were shifted by atmospheric winds to more remote locations. Most swarms however, are almost buried in strata of clay and silt. The Peterborough swarm has been described as a "carpet" lying directly on Paleozoic bedrock. (63) (Illustration 3, above) It is highly unlikely that atmospheric winds could be the main separator of debris. Mountain ranges are composed of debris too massive to be greatly influenced by atmospheric effects, yet they display the same adjacent regime of fossil-rich strata.

However, the physical processes ejecting debris into comet tails could easily be responsible for the dual regimes found in esker/drumlin swarms. Donnelly's observations of the iron rich and iron poor clays are an example of the transformation that must occur in the electromagnetic medium of a comet's tail.

**The Frozen Landscape**

Let's consider a typical esker/drumlin swarm such as the Peterborough 'esker' prior to its full mobilization and deposition by the returning seasons. We assume that the ridge structure was wholly buried beneath snow and ice, while the lighter organic and inorganic materials were diffused throughout a wider belt extending a few kilometers on either side of the main ridge formations. When the global upheaval accompanying this event had run its course and the sky cleared sufficiently to re-create the normal cycle of seasons, the ice sheet would begin to retreat. This process revealed the pre-urban landscape we see today. Before the rains and summer melting began to cause major recession, the buildup of ice and season snows would have turned present valleys into slowly-moving rivers of ice. Swarms straddling these valleys would be broken away, their debris pushed into the major ice routes such as the Hudson and St. Lawrence River troughs.

**The Run Off**

Moraines are only the more obvious ridges left by the retreating ice. Complex hydrocarbons and simple salts also left their imprint on the landscape.

**The Dam Ridges**

When not torn away by valley glaciers, esker/drumlin swarms acted as effective dams and sluices. Their greater bulk would have assured their earthward movement at a much greater rate than the lighter and more scattered silts and clays. The subsequent run-off, carrying the slurries, suspensions and solutions trapped in the ice sheet, would find the ridges and as water will do, take the easiest path through or over the obstacle. The slurries would tend to settle against the upslope side of the swarm ridges, leaving their characteristic layered imprint (varves) built up, sometimes in annual cycles, against the upslope side of the ridges.

The balance of these muddy rivers would find their way through the gaps in the swarm, filling their interstices and creating plateaux.

**Concentric Arcs of Hydrocarbons**

Hydrocarbons combined with the outwash and turned into a gelatinous mass. (64) Less dense than water, these sticky masses would float on the run-off's surface and be lifted above or passed through the swarm ridges and deposited at the base of the retreating ice sheet. These deposits, conforming to the lobes of retreating ice and snow, would form concentric arcs of hydrocarbons, with each arc representing a single season.

These arcs reveal the same dynamic behavior and basic shape as moraines. They may still be seen in northern continental terrains where they are called 'strings'. Whether collected in concentric strings or in basins, huge deposits of this complex hydrocarbon, called peat, are found on all continents previously covered by the last ice sheet. The peat deposits correspond to the great coal deposits found in the gaps between mountain ranges such as the coal fields found in the northeast of the United States. (Illustration 17, above)

# CONCLUSION

**Tantalizingly Close**

Donnelly recognized the drift as cometary debris, but in his rejection of the reigning glacial hypothesis, failed to recognize its true association with ice. Ehlers, in a recent text, summarizes the early attitude:

"The origin of tills in formerly-glaciated areas has been a matter of debate since the emergence of glacial theory. In North Germany in the late nineteenth century, tills were first regarded as super glacial morainic deposits of the Pleistocene ice sheets. The concept was only refuted after Nansen (Norwegian explorer Fridtjof Nansen) had crossed the Greenland ice sheet. Nowhere on the inland ice did he and his expedition find any super glacial debris." (65)

One is tempted to imagine a chance encounter between Donnelly and Nansen, with his now-defeated conviction that somehow drift was levitated to the top of ice sheets. Donnelly's catastrophic scenario cast ice not as a direct contributor but a byproduct of the lash of the great comet's tail. He felt that the drift struck with such ferocity that the resulting heat created earth-enfolding clouds of condensing vapor, which then fell as the deep snows of the great winters of 'Klimasturz' of folklore memory. (66) Nansen, presumably a mainstream naturalist who would be inimical to catastrophic concepts such as the deluge of fundamentalist Christian doctrine and the ice-buried earth of Aggissiz, would have probably given the wilder epic of Donnelly little consideration.

**The Unfinished Puzzle**

And so catastrophic theory has remained with scientists still looking for the connecting pieces of a jig-saw puzzle that will not quite go together. Despite the turn to the study of catastrophism the earth sciences have taken in the last decade, we are still presented with a bleak and distorted portrait of our planet's evolution. New life still slowly emerges, 'red in tooth and claw' from the smoking ruins of an atmosphere poisoned by a death star or rogue comet. Yet paleontologists agree that we have not only evolved but proliferated after each of these earth ages. Greater numbers of species appear to radiate from these global crises than expire in their physical fury and they appear to do so in co-operation with their fellow fauna and flora. (67)

Do comets have a benign as well as a destructive role to play in these evolutionary epics? Because they only cover the earth partially, existing and emerging flora and fauna are not completely destroyed. On the contrary, their tails may contain the hydrocarbons and other complex molecules needed to renew a flagging biosphere and provide new physical environments for emerging life forms.

## END NOTES

1.  With the exception of volcanically formed mountains.

2.  Ehlers, Quaternary and Glacial Geology, Toronto, 1996, p. 65., and I. J. Smalley and D. J. Unwin, "The Formation and Shape of Drumlins and their Distribution and Orientation in Drumlin Fields", Journal of Glaciology vol. 7 pp. 388- 394, 1968

3.  See Peterborough drumlin/esker swarm illustration (illus. 3)

4.  Ibid.

5.  Ibid.

6.  I. Banerjee and B. C. McDonald, "Nature of Esker Sedimentation in Glaciofluvial and Glaciolacustrine Sedimentation". Editor A. V. Jopling and B. C. McDonald, Society of Economic Paleontologists and Mineralogists. Special publication 23, pp. 132-154, 1975.

7.  See super-swarm illustration. R.F. Flint, Glacial and Pleistocene Geology, New York, 1957. p.155.

8.  Ehlers, Op. cit., p.92.

9.  Jopling, Op. cit., "Shaler (N.S. Shaler, 1884), on the origin of kames, Boston Soicety of Natural History proceedings, vol. 23, pp. 36-44, mentioned that in the New England area, the eskers, which he had provisionally referred to as serpent kames, were commonly attributed to the work of Indians, hence the local terminology 'Indian ridges'." p.134.

10. J. K. Charlesworth, The Quaternary Era, London. 2 volumes, p. 423, "In Denmark they (eskers) were associated with a goblin with a leaky sandbag."

glacial deposits in the geologic record be the products of debris flows caused by large impact events (from space)?

## ARTICLES

# GLOSSARY

**Core**            See ridge.

**Deluvialist(s)**  Geologists who believe the surface features of the earth were shaped by Noah's deluge.

**Divide**          An area of land or water marking the boundary from which the glacial ice sheets of the last ice age apparently radiated outward.

**Drift**           A general term for all rock material supposed to have been transported by glaciers and deposited directly from the ice or through the agency of meltwater. It is generally applied to Pleistocene deposits in large regions that no longer contain glaciers.

**Drumlin**         A ridge of sand and gravel left behind when the continental glacial ice of the last age melted. It ranges in length from 400 to 2000 meters (average length is 500 meters) and from 8 to 60 meters in height (average height is 30 meters). It is usually shorter in length than an esker (see below) and has a blunt nose pointing in the direction from which the ice supposedly approached and has a gentler slope tapering in the other direction.

**Debris tail**     The tail of a comet composed of sand, gravel and other (millimeter-sized or larger) cometary debris.

**Esker**           A ridge of sand and gravel left behind when the continental glacial ice of the last age melted. It ranges in length from less than 100 meters to more than 500 kilometers (counting gaps) and from 3 to more than 300 meters in height. Both ends of the ridge are blunt.

**Fluvial**         Produced by the action of a stream.

**Gas tail**    The number one tail of a comet composed of gas and pointing directly away from the sun.

**Glacialist(s)**   Geologists who believe the surface features of the earth were shaped by the vast ice sheets of the last ice age.

**Jet**      A ribbon or filament composed of sand, gravel and other cometary material ejected in a stream from discrete areas of a comet's nucleus.

**Kame**     A mound, knob or short irregular ridge composed of stratified sand and gravel, deposited at the margin of a melting glacier.

**Klimasturz**   A catastrophic change of climate, during which, under mysterious circumstances, all the mammoths of Siberia perished.

**Mantle**    The sedimentary regime covering esker/drumlin swarms.

**Morarne**    A ridge of stratified glacial drift, chiefly till, deposited by the direct action of a static or retreating continental ice sheet at its perimeter.

**Ridge**     A general term for a long narrow core with steep sides, exclusive of the mantle.

**Stria**      One of a series of fine grooved lines or threads on the surface of a rock.

**Strings**    Morainic deposits of peat.

**Swarm**    A group of roughly parallel eskers, drumlins and related ridges.

**Till**      Unstratified drift, deposited directly by a glacier without reworking by meltwater.

**Varve**     A sedimentary sequence of laminae deposited in a body of still water within one year's time.